\documentclass[twocolumn, tightenlines, amsmath, amssymb, superscriptaddress, table,xcdraw]{revtex4-1}
\usepackage{graphicx} 
\usepackage{fourier}
\usepackage{physics}
\usepackage[
separate-uncertainty=true, 
multi-part-units=single,]
{siunitx}
\usepackage{hyperref}
\usepackage{amsmath}  
\usepackage{amsthm} 
\usepackage{dsfont}
\usepackage{booktabs} 
\usepackage{changes}
\usepackage[]{xcolor}
\usepackage{xspace}
\usepackage{array}
\usepackage[linesnumbered, ruled, vlined]{algorithm2e}

\newtheorem{theorem}{Theorem}

\newcommand{\hvec}[1]{\hat{\vec{#1}}}  
\newcommand{\TVDTEST}{\text{TVD}_\text{test}}
\newcommand{\FA}{\mathcal{F}_a}
\newcommand{\CTWORES}{C_2^\text{(res)}}
\newcommand{\CTWORESHAT}{\hat{C}_2^\text{(res)}}
\newcommand{\CTWOEXT}{C_2^\text{(ext)}}
\newcommand{\CTWOEXTHAT}{\hat{C}_2^\text{(ext)}}
\newcommand{\CTWORESEXT}{C_2^\text{(res/ext)}}
\newcommand{\CTWORESEXTHAT}{\hat{C}_2^\text{(res/ext)}}
\newcommand{\CZEROEXT}{\vec{c}_0^\text{(ext)}}

\newcommand{\CZERORES}{\vec{c}_0^\text{(res)}}
\newcommand{\PHIRES}{\vec{\phi}^\text{(res)}}
\newcommand{\PHIEXT}{\vec{\phi}^\text{(ext)}}

\newcommand{\NPS}{n_\text{PS}}
\newcommand{\NCPS}{n_\text{CPS}}
\newcommand{\deltacj}{\ensuremath{\Delta c_j}}

\begin{document}

\title{
Resource-efficient crosstalk mitigation for the high-fidelity operation of photonic integrated circuits with induced phase shifters 
}

\author{Andreas Fyrillas}
\affiliation{Quandela, 7 Rue Léonard de Vinci, 91300 Massy, France}
\affiliation{Centre for Nanosciences and Nanotechnologies, CNRS, Université Paris-Saclay, UMR 9001, 10 Boulevard Thomas Gobert, 91120, Palaiseau, France}
\author{Nicolas Heurtel}
\affiliation{Quandela, 7 Rue Léonard de Vinci, 91300 Massy, France}
\affiliation{Université Paris-Saclay, CNRS, ENS Paris-Saclay, Inria, Laboratoire Méthodes Formelles, 91190,
Gif-sur-Yvette, France}
\author{Simone Piacentini}
\author{Nicolas Maring}
\author{Jean Senellart}
\affiliation{Quandela, 7 Rue Léonard de Vinci, 91300 Massy, France}
\author{Nadia Belabas}
\affiliation{Centre for Nanosciences and Nanotechnologies, CNRS, Université Paris-Saclay, UMR 9001, 10 Boulevard Thomas Gobert, 91120, Palaiseau, France}

\begin{abstract}
\hrule
\vspace{0.5cm}
Photonic integrated circuits (PICs) are key platforms for the compact and stable manipulation of classical and quantum light. Imperfections arising from fabrication constraints, tolerances, and operation wavelength limit the accuracy of intended operations on light and impede the practical utility of current PICs. In particular, crosstalk between reconfigurable phase shifters is challenging to characterize due to the large number of parameters to estimate and the difficulty in isolating individual parameters. Previous studies have attempted to model crosstalk solely as an interaction between controlled phase shifters, overlooking the broader scope of this issue. We introduce the concept of induced phase shifter, arising from crosstalk on bare waveguide sections as predicted by simulations, resulting in an exhaustive description and systematic analysis of crosstalk. We characterize induced phase shifters in physical devices using a machine learning-based method and propose a mitigation framework. This framework further allows to establish a criterion certifying that a given interferometer has a sufficient number of degrees of freedom adequately laid out to fully mitigate crosstalk. Our approach is experimentally validated on a 12-mode Clements interferometer. We demonstrate the efficacy of our extended crosstalk model to accurately recover physical crosstalk properties of the PIC and cancel induced phase shifters following our mitigation framework.

\vspace{0.5cm}
\hrule
\end{abstract}

\maketitle

\section{Introduction}

Photonic integrated circuits (PICs) embark optical components in a tightly integrated platform for light manipulation with enhanced compactness, scalability and stability. PICs are essential both for classical \cite{Bogaerts2020} and quantum optics \cite{Wang2020}, with applications ranging from microwave photonics \cite{Marpaung2019}, optical beamforming \cite{Heck_2017}, and high-precision sensing \cite{Luan2018} to quantum computing \cite{Zhong2020, Maring2024}, quantum communication \cite{Luo2023}, quantum cryptography \cite{Fyrillas2024a}, and quantum sensing \cite{Polino2019}. We consider here programmable PICs for light manipulation, featuring directional couplers, and reconfigurable phase shifters (PSs). Directional couplers act as beamsplitters of fixed reflectivity, while PSs, controlled  by voltage or electric current, enable the PIC to perform a wide variety of operations on input light states. Reconfigurable PSs harness thermo-optic effects \cite{Harris2014}, strain-induced birefringence \cite{Dong2022} or electro-optic effects \cite{Li2020} to apply phase shifts on the guided light. 

In general, crosstalk between PSs is an expected consequence of their compact integration. Our focus is here on voltage-controlled thermo-optic PSs and thermal crosstalk, which is the most widely used PS technology. Thermo-optic PSs generate heat and leverage the temperature dependence of the refractive index. The generated heat then diffuses within the PIC and manifests as crosstalk between the circuit's PSs.  Crosstalk between PSs is recognized as a major imperfection in the integrated photonics litterature \cite{Banerjee2023}. The mitigation of crosstalk presents a considerable challenge due to the multitude of contributing parameters and the complexity of interferometer meshes, which hinder the isolation and control of individual crosstalk components. The formalism used in \cite{Bandyopadhyay2022} to account for thermal crosstalk was first introduced in \cite{Milanizadeh2019, Milanizadeh2020} as a matrix relation between the target phases and the actual implemented phase shifts on each PS. A similar matrix-based approach is used in \cite{Fyrillas2024a, Pont2024, Pentangelo2024, Fyrillas2024b, Rodari2024} to relate the implemented phase shifts to the applied voltages or electric currents. In particular, our previous work \cite{Fyrillas2024b} achieved the highest recorded fidelity for implemented unitary matrices by using a machine-learning assisted method to retrieve the phase-voltage relation parameters of a physical PIC. The recovered phase-voltage relation exhibits unphysical long-range crosstalk. This hints towards a mechanism overlooked by current crosstalk models, and suggests that a physically accurate and systematic crosstalk description for PICs is still missing. 
In this work,
\begin{itemize}
    \item we extend existing crosstalk descriptions by introducing an effective optical component accounting for parasitic phase shifts on bare waveguide sections, confirmed in simulations (Section \ref{sec:induced_ps}).
    \item We provide and benchmark a machine learning based-method for measuring crosstalk in PICs in the extended crosstalk paradigm (Section \ref{sec:charac}).
    \item We mitigate crosstalk in the extended crosstalk framework, which demonstrates that mitigation is fundamentally infeasible in certain interferometer meshes. Based on this new understanding, effectively mitigating crosstalk through optimized PIC designs is key to improving control, scalability, and fidelity, while minimizing the number of on-chip components. We establish a criterion for certifying the inherent ability of a given interferometer to cancel crosstalk (Section \ref{sec:mitigation}).
    \item Finally, we experimentally assess the improvements achieved thanks to the extended crosstalk model (Section \ref{sec:validation}).
\end{itemize}

\section{Crosstalk model extension with induced phase shifters}
\label{sec:induced_ps}

\begin{figure*}[ht]
    \centering
    \includegraphics[width=\linewidth]{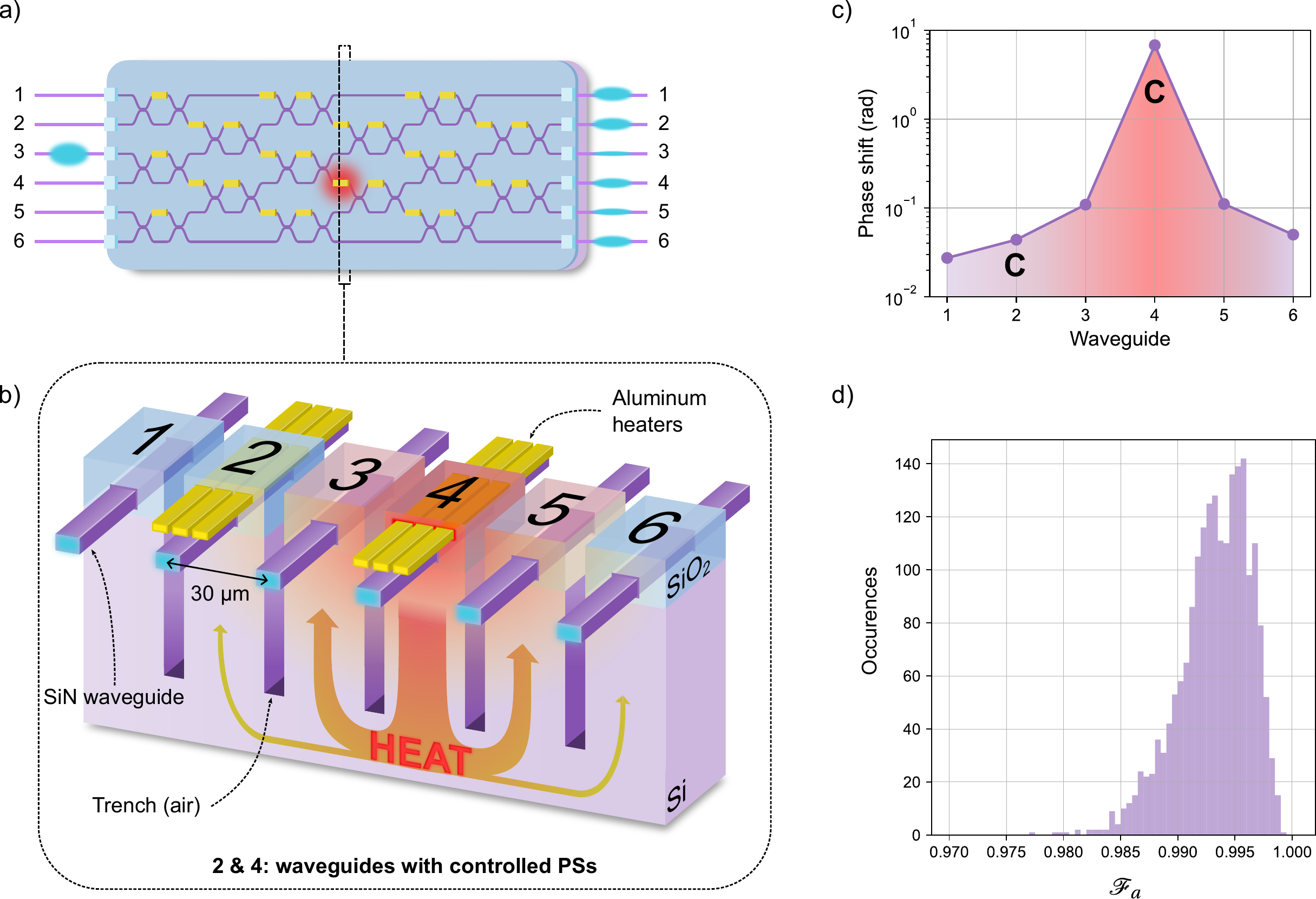}
    \caption{\textbf{Thermal crosstalk in photonic integrated circuits induces parasitic phase shifts in neighboring waveguides.}
    \textbf{a)} 6-mode Clements interferometer with thermo-optic phase shifters (PSs). Yellow rectangles indicate the location of controlled phase shifters. Red area indicates an active controlled PS generating heat that induces a target phase shift on light propagating through its associated waveguide portion. 
    \textbf{b)} Cross section of photonic circuit (not to scale, see App.~\ref{app:th_sim}). The controlled PS on waveguide 4 is actuated and generates heat which diffuses. Blue: silicon dioxide (SiO$_2$) layer. Dark purple: silicon nitride (SiN) waveguides. Light purple: silicon (Si) substrate. Yellow: aluminum heaters. The trenches are etched in the Si substrate and filled with air. Only waveguides 2 and 4 feature a controlled PS (the controlled PS on waveguide 2 is not actuated here).
    \textbf{c)} Simulated values of induced phase shifts in each waveguide of the cross section (see Methods). The letter "C" indicates the waveguides featuring a controlled PS. The applied bias voltage on the actuated controlled PS of waveguide 4 is 14 V, which is the value inducing a $2\pi$ phase shift on waveguide 4 with realistic parameters (see Methods).
    \textbf{d)} Histogram of amplitude fidelities between target unitary matrices and implemented matrices on PIC. Only crosstalk between controlled PSs is taken into account when solving the phase-voltage relation of the PIC (see Methods "Amplitude fidelity measurement"). The average amplitude fidelity is $\SI{99.3}{\%}$. 
    }
    \label{fig:cross_section}
\end{figure*}

Given a PIC, let $n_\text{CPS}$ be the number of \textit{controlled} PSs, that is the number of phase shifters that are physically fabricated as such on the PIC and directly controlled by voltage. PIC crosstalk models usually describe crosstalk as a mutual influence between controlled PSs \cite{Milanizadeh2019, Bandyopadhyay2022, Fyrillas2024b, Zhu2020, Banerjee2023}. This influence is captured by a matrix relation relating the applied voltages $\vec{V}$ to the implemented phase shifts $\PHIRES$ restricted to controlled PSs. Hence, the phase-voltage relation of a PIC in what we call the \textit{restricted crosstalk model} has the general form
\begin{equation}
    \PHIRES = \sum_{k \neq 0} C_k^\text{(res)} \cdot \vec{V}^{\odot k} + \CZERORES 
\label{eq:phase_voltage_restricted}
\end{equation}
where $C_k^\text{(res)}$ are crosstalk matrices of size $n_\text{CPS}\times n_\text{CPS}$, $^{\odot}$ is element-wise exponentiation and $\vec{c}_0^\text{(res)}$ is the vector containing the $n_\text{CPS}$ passive phases. The diagonal elements of a crosstalk matrix $C_k^\text{(res)}$ correspond to the \textit{self-heating} coefficients. Self-heating coefficients quantify the efficiency with which PS heaters modify the phase of light propagating in the waveguide they are located on. Off-diagonal elements of crosstalk matrices account for crosstalk between PSs. For ideal thermo-optic phase shifters, only the $k=2$ term remains in Eq.~\ref{eq:phase_voltage_restricted}. We will consider for simplicity PICs with phase-voltage relations of the form $\PHIRES = \CTWORES \cdot \vec{V}^{\odot 2} + \CZERORES$. 

Restricted crosstalk models do, however, not reflect the physical reality of actual crosstalk. Physical crosstalk, related to heat diffusion for thermo-optic PSs, is indeed expected to also create phase shifts on waveguide portions that do not feature any controlled PS. Analogous effects may also exist for mechanical and electro-optic phase shifters, by spread of strain or stray electric fields to neighboring waveguides. To demonstrate this effect for thermo-optic PSs, we perform numerical simulations on a silicon nitride (SiN) PIC featuring 6 \textit{modes} (number of waveguides running through the PIC), shown on Fig.~\ref{fig:cross_section}a. The impact of crosstalk is assessed by computing the generated phase shifts in waveguides neighboring an active controlled PS (see Methods and App.~\ref{app:th_sim} for details on the simulations). The PIC, consistent with typical designs, is fitted with trenches to increase phase shifter efficiency and reduce heat diffusion in the circuit \cite{Alemany2021, Ceccarelli2020} (see cross section on Fig.~\ref{fig:cross_section}b). The resulting phase shifts displayed on Fig.~\ref{fig:cross_section}c confirm that waveguides that do not initially feature a PS heater acquire an unwanted phase shift due to heat diffusion. This effect is not included in the restricted crosstalk model, which by definition only accounts for crosstalk between controlled PSs. 

We show on Fig.~\ref{fig:cross_section}d that the presence of induced PSs significantly impacts interferometer behavior. To evaluate this, we implement a set of target unitary matrices on the simulated PIC featuring induced PSs. Each matrix encodes the action of the interferometer on light \cite{Reck1994}. Following the procedure in Methods ("Amplitude fidelity measurement"), the target matrices are implemented considering only crosstalk between controlled PSs to compute the applied voltages. The resulting average amplitude fidelity between targeted and implemented matrices is $\SI{99.3}{\%}$ due to the presence of induced PSs. The amplitude fidelity between two unitary matrices $U$ and $V$ is defined as \cite{Pentangelo2024}
\begin{equation}
    \FA(U,V) = \frac{\Tr(|U|^T\cdot |V|)}{m}
\label{eq:fidelity}
\end{equation}
where the absolute value is applied element-wise and $m$ is the number of modes. For comparison, the highest experimentally achieved average amplitude fidelity between implemented and target Haar-random unitary matrices is $\SI{99.7}{\%}$ (in \cite{Fyrillas2024b} on a 12-mode Clements interferometer).

\begin{figure*}[ht]
    \centering
    \includegraphics[width=\linewidth]{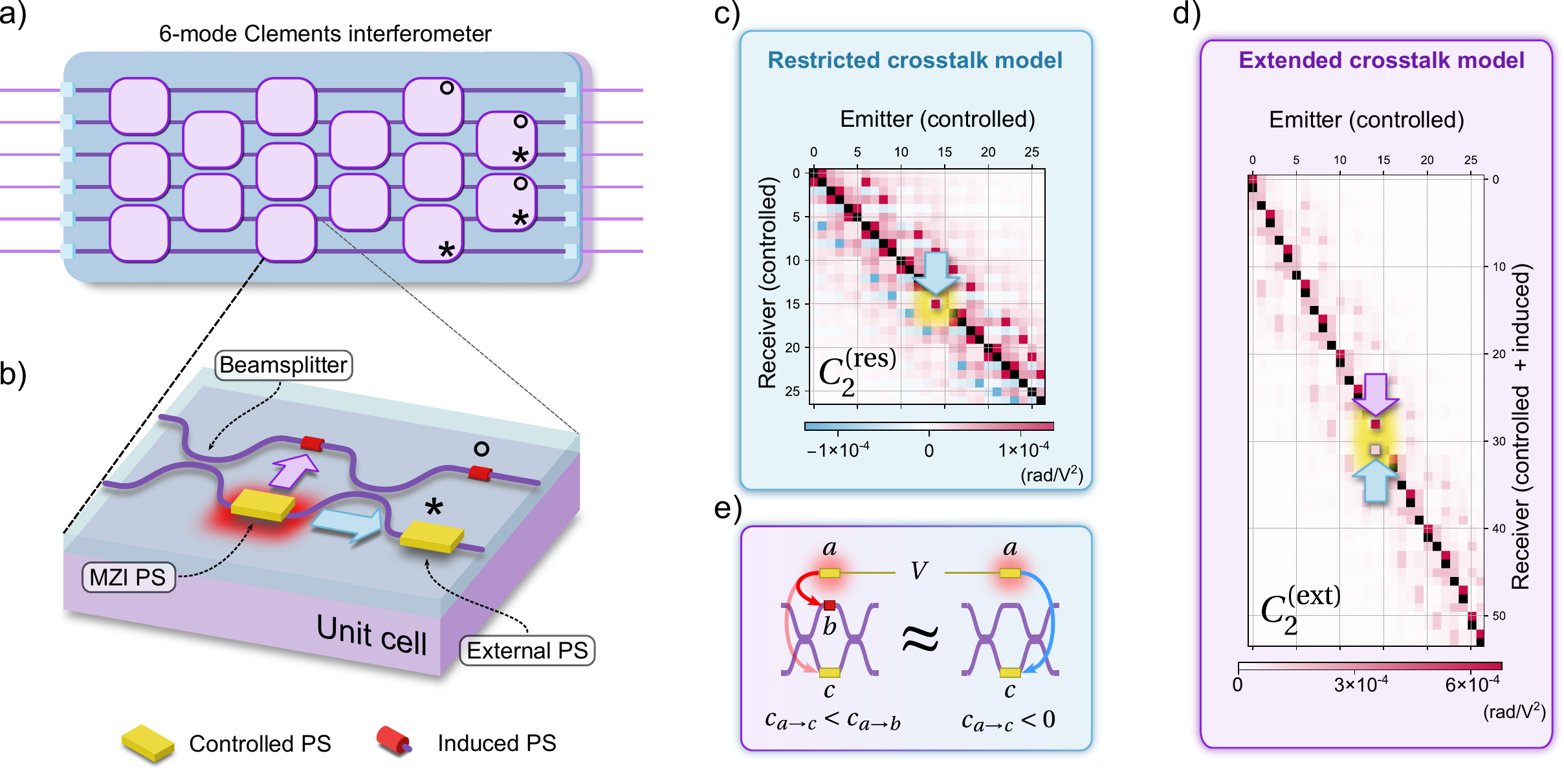}
    \caption{
    \textbf{Extended crosstalk model taking into account induced phase shifters.}
    \textbf{a)} Schematic of 6-mode Clements interferometer. The basic unit cell \textbf{b)} of the interferometer contains two beamsplitters, two controlled phase shifters (PSs): one Mach-Zehnder interferometer (MZI) PS and one external PS. The unit cell also features two induced PSs. "*" or "°" symbols on unit cells in a) indicate that the unit cell lacks PSs marked with the same symbols on b), as they do not produce any measurable effect on light intensity detectors. Overall, the PIC has $\NPS=54$ PSs, among which $\NCPS=27$ are controlled.
    \textbf{c)} Simulated restricted crosstalk matrix. The restricted crosstalk model only considers interactions between controlled PSs via a square crosstalk matrix $\CTWORES$ in the restricted phase-voltage relation (Eq.~\ref{eq:phase_voltage_restricted}). The $\CTWORES$ matrix displayed here is simulated from the circuit layout (see Methods, "Simulation benchmark of the training process"). The blue arrow on $\CTWORES$ points at the coefficient corresponding to the influence symbolized by the blue arrow on the unit cell b). Diagonally dominant coefficients are masked.
    \textbf{d)} Simulated extended crosstalk matrix. The extended crosstalk model includes induced PSs representing phase shifts appearing due to crosstalk on waveguides devoid of any controlled PS. The associated simulated extended crosstalk matrix $\CTWOEXT$ is displayed, along with two colored arrows pointing at the coefficients corresponding to the influence symbolized by the color-coded arrows on b). 
    \textbf{e)} Configuration leading to negative coefficients in the restricted crosstalk matrix. $c_{x \rightarrow y}$ denotes the crosstalk coefficient associated to the influence of PS $x$ on PS $y$. Left: extended crosstalk model. A voltage $V$ is applied on the heat emitter $a$. The MZI has an induced PS $b$ on the top arm and a controlled PS $c$ on the bottom arm. PS $b$ is closer to $a$, thus $0 < c_{a \rightarrow b} < c_{a\rightarrow c}$. As a result, the phase difference between the two arms $\phi_c - \phi_b = (c_{a\rightarrow c} - c_{a\rightarrow b})V^2 < 0$ decreases with increasing voltage $V$. In addition, the unitary matrix of the MZI acquires a global phase $e^{i(c_{a\rightarrow c} + c_{a\rightarrow b})V^2}$. The impact of crosstalk on the phase difference can be taken into account in the restricted crosstalk model (right), in which the MZIs only have a controlled PS, with $c_{a\rightarrow c}<0$. The impact on the global phase is however not captured by the restricted crosstalk model.
    }
    \label{fig:model_extension}
\end{figure*}

To account for these parasitic induced phase shifts, we propose to extend the standard restricted crosstalk models by introducing \textit{induced} PSs in the circuit. Induced PSs are not directly controlled by voltage, but arise from the refraction index change caused by heat diffusion in bare waveguide sections. We thus consider in the extended crosstalk model that all the circuit waveguide portions feature either an initially present controlled PS, or an added induced PS. This is shown on Fig.~\ref{fig:model_extension}b. As a result, the restricted square crosstalk matrix $\CTWORES$ of size $\NCPS\times\NCPS$ (see Fig.~\ref{fig:model_extension}c) is augmented to an extended rectangular matrix $\CTWOEXT$ of size $\NPS\times\NCPS$ (see Fig.~\ref{fig:model_extension}d) with $\NPS$ the total number of on-chip PSs (including controlled and induced ones). 

Note that extended crosstalk matrices contain only positive coefficients, because crosstalk increases the phase of each PS (assuming that the refractive index increases with temperature, as is the case for silica glass or SiN). In the restricted model, controlled PSs in Mach-Zehnder interferometers (MZI) can have a negative crosstalk coefficient due to the situation depicted in Fig.~\ref{fig:model_extension}e. 

Similarly, the vector of implemented phases $\PHIRES$ and the passive phase vector $\CZERORES$ are promoted from $\NCPS$ to $\NPS$ components in their extended versions $\PHIEXT$ and $\CZEROEXT$ which include controlled and induced PSs. We set the passive phase of induced PSs to zero by convention, such that the restricted and extended crosstalk models yield the same PIC state when no voltage is applied. The extended phase-voltage relation reads
\begin{equation}
    \PHIEXT = \CTWOEXT \cdot \vec{V}^{\odot2} + \CZEROEXT.
\label{eq:phase_voltage_extended}
\end{equation}

\section{Characterization of crosstalk using machine learning}
\label{sec:charac}

We discuss in this section a method for experimentally measuring extended crosstalk matrices $\CTWOEXT$ of PICs, based on the machine learning-assisted method we introduced in \cite{Fyrillas2024b}. The method adopts a clear-box approach to PIC characterization. In the clear-box approach, the machine learning model (MLM) acting as the virtual replica of the PIC to characterize is built on a physical PIC model. This is in contrast with methods using virtual replicas based on black-box neural networks \cite{Cimini2021, Youssry2024}. The clear-box MLM is trained to replicate the behavior of the hardware PIC to characterize. This requires a training and a test dataset of samples acquired on the PIC. For each sample, light is injected into a PIC input port $i$, a list of random voltages $\vec{V}$ is applied on the PIC and the resulting light intensity output distribution $\vec{p}$ is measured (normalized to sum to 1), yielding a sample $(\vec{V}, i, \vec{p})$. The restricted MLM of \cite{Fyrillas2024b} is extended by adding induced PSs to the machine learning model. As described in Sec. \ref{sec:induced_ps}, induced PSs are not directly controlled by a voltage in the model, but emerge from a realistic crosstalk that is described by an rectangular crosstalk matrix $\CTWOEXT$.

\begin{figure*}[ht]
    \centering
    \includegraphics[width=\linewidth]{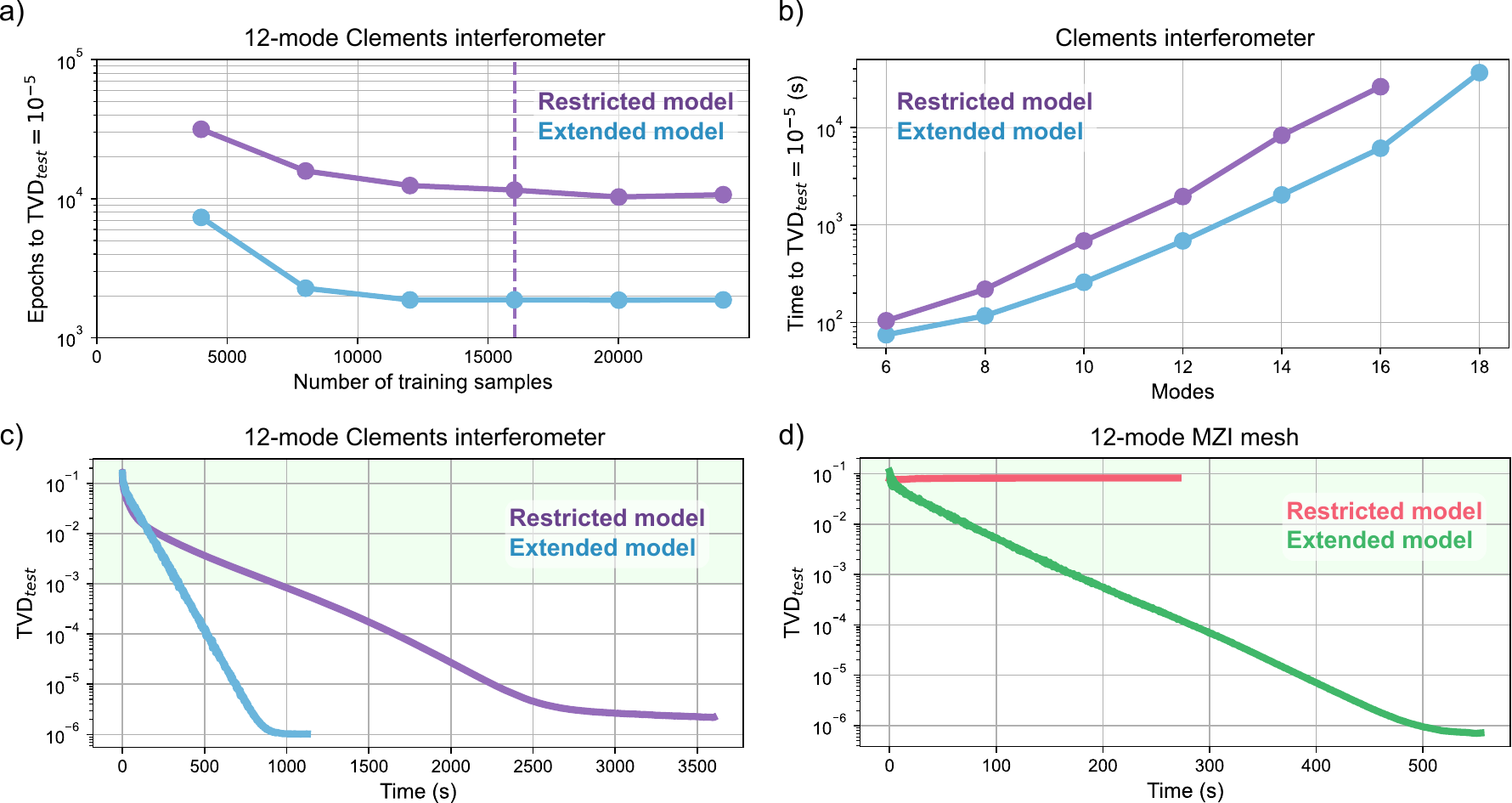}
    \caption{
    \textbf{Simulation benchmarking the characterization of induced phase shifters.}
    The simulations are performed on simulated 12-mode photonic integrated circuits (PICs) (see Methods) featuring induced phase shifters (PSs). We compare the training performance of machine learning models (MLMs) equipped with restricted and extended crosstalk models.
    \textbf{a)} Number of epochs needed to reach $\text{TVD}_\text{test}=10^{-5}$ versus the number of training samples used for training the restricted and extended crosstalk models on a 12-mode Clements interferometer. The purple dashed line indicates the number of samples corresponding to a data-to-parameter ratio of 1 for the restricted model. From here on, to train the restricted and extended models, we always use a number of training samples corresponding to a data-to-parameter ratio of 1 for the restricted model
    \textbf{b)} Elapsed time needed for the restricted and extended MLMs to reach $\text{TVD}_\text{test}=10^{-5}$ for Clements interferometers with increasing numbers of modes. 
    \textbf{c)} TVD$_\text{test}$ as a function of elapsed training time in the case of a 12-mode Clements interferometer for restricted and extended MLMs. The number of training samples in both cases is 16 000 (data-to-parameter ratio equal to 1 for the restricted model). Green area indicates experimentally accessible values of TVD$_\text{test}$.
    \textbf{d)} Same as c) for a mesh of MZIs, which is a Clements interferometer whose external controlled PSs (see Fig. \ref{fig:model_extension}b) have been converted into induced PSs ($\NPS=252$ PSs, among which $\NCPS=66$ are controlled). Both models were trained with $\approx 4 000$ samples (data-to-parameter ratio equal to 1 for the restricted model).
    }
    \label{fig:convergence}
\end{figure*}

We compare the ability of the MLMs equipped with restricted and extended crosstalk models to learn the behavior of crosstalk in a hardware PIC. The hardware device to characterize is here replaced by a simulated idealized PIC whose only imperfection is crosstalk, i.e. all beamsplitters are symmetric, the PIC is perfectly transmissive but the crosstalk matrices $\CTWORES$ and $\CTWOEXT$ have nonzero off-diagonal elements. The simulated PIC has an extended crosstalk model whose parameters are unknown to the MLMs (see Methods). During the training process, the MLMs attempt to converge to adequate parameters for their crosstalk matrix such that the mean square error (MSE) between the training dataset light intensity distributions $\vec{p}$ and the model predictions $\hvec{p}$ is minimized (quantities with a hat represent MLM predictions). In practice, the estimated crosstalk matrix $\CTWORESHAT$ (resp. $\CTWOEXTHAT$) of the restricted (resp. extended) MLM is tuned via gradient-descent (see Methods), where each step is denoted \textit{epoch}. This corresponds to the machine learning stages denoted "ML" in Figure 2 of \cite{Fyrillas2024b}.

The distance between a dataset light intensity distribution $\vec{p}$ and its corresponding prediction $\hvec{p}$ is quantified by the total variation distance (TVD):
\begin{equation}
    \text{TVD} \left(\vec{p}, \hvec{p} \right) = \frac{1}{2}
    \sum_i |p_i - \hat{p}_i|
\end{equation}
where $\text{TVD} = 0$ means $\vec{p}=\hvec{p}$ and $\text{TVD} = 1$ signifies that $\vec{p}$ and $\hvec{p}$ are maximally different. The prediction accuracy of an MLM is gauged by the average TVD on the test dataset and denoted $\TVDTEST$.

\paragraph{Benchmark on a Clements interferometer and training sample requirement} 
The restricted and extended MLMs are trained on a 12-mode Clements interferometer \cite{Clements2016} ($\NPS=252$ PSs, among which $\NCPS=126$ are controlled) with an increasing number of training samples. The number of parameters to train in the restricted (resp. extended) crosstalk model is the number of matrix elements in $\CTWORES$ (resp. $\CTWOEXT$), that is $\NCPS\times\NCPS \approx 16 000$ (resp. $\NCPS\times\NPS\approx32 000$). Fig.~\ref{fig:convergence}a reveals that the number of training samples required to train the extended MLM does not increase compared to the restricted MLM. The invariance of the \textit{data-to-parameter ratio} (number of training samples / number of MLM parameters) is especially relevant for practical aspects of characterization. Indeed the temporal bottleneck in the characterization protocol is the collection of the training and test datasets \cite{Fyrillas2024b}, which are themselves limited by the PIC reconfiguration and detector integration times (see Methods). In the following, we always train the restricted and extended MLMs with a number of training samples corresponding to a data-to-parameter ratio of 1 for the restricted model. This corresponds to 16 000 training samples for the 12-mode Clements interferometer. 

Interestingly, the extended MLM requires significantly fewer gradient descent epochs to reach a target $\TVDTEST$ value, which we attribute in part to the doubled amount of degrees of freedom granted by the extended crosstalk matrix $\CTWOEXTHAT$. Consequently, each epoch is longer to compute for the extended MLM ($\SI{302}{ms}$ against $\SI{235}{ms}$ for the restricted MLM). There is nevertheless a substantial net time gain when training the extended MLM, as shown on Fig.~\ref{fig:convergence}b. Fig.~\ref{fig:convergence}c displays the evolution of $\TVDTEST$ during training for the restricted and extended MLMs. 

When using single-photon detectors, the count rate measurements are affected by Poisson shot noise, which limits $\TVDTEST$ to values above 0.1 \% as demonstrated in \cite{Fyrillas2024b}. Empirically, the $\TVDTEST$ is similar when performing the characterization with a continuous-wave laser and powermeters. Hence, the advantage of the extended model over the restricted model in terms of attained $\TVDTEST$ values is not experimentally relevant in our case. 

\paragraph{Benchmark on an MZI mesh} 
We repeat the simulation of the previous paragraph with an interferometer featuring only the MZIs of a 12-mode Clements interferometer. The external PSs of the Clements mesh (see Fig. \ref{fig:model_extension}b) are replaced by induced PSs in the MZI mesh. We observe on \ref{fig:convergence}d that the restricted model does not converge. On the contrary, the extended model manages to replicate the PIC behavior. Hence, the extended MLM converges on more interferometer meshes, featuring induced PSs, compared to the restricted model. This observation is justified in Section \ref{subsec:criteria}. 

We verify that the the inability of the restricted model to converge is indeed due to the presence of realistic induced PSs and not tied to the particular MZI mesh chosen here. Indeed, removing the induced PSs from the simulated MZI mesh allows the restricted model to converge.

\section{Crosstalk mitigation in the extended framework}
\label{sec:mitigation}

It is assumed in this section that the crosstalk matrix $\CTWORESEXT$ of the hardware device is known. In practice, $\CTWORESEXT$ is approximated by the crosstalk matrix $\CTWORESEXTHAT$ estimated from characterization by the restricted/extended MLM of Section~\ref{sec:charac}. Mitigating crosstalk consists in solving the phase-voltage relation (restricted Eq.~\ref{eq:phase_voltage_restricted} or extended Eq.~\ref{eq:phase_voltage_extended}) of the device to find a set of voltages achieving the specified target phases. This requires the inverse of the crosstalk matrix (see App.~\ref{app:solver} for a description of our solver). Square crosstalk matrices $\CTWORES$ are usually invertible by diagonal dominance. However, in extended crosstalk models, the crosstalk matrix $\CTWOEXT$ is rectangular (see Section \ref{sec:induced_ps}), thus the inverse is not defined. Specifically, $\CTWOEXT$ has more rows than columns, which entails that the phase-voltage relation is an underdetermined system of equations. 

In the rectangular case, the algebraic matrix inverse may be replaced by the Moore-Penrose pseudo-inverse. The induced phase shifts are set to zero in the target phases vector $\PHIEXT$ given to the solver, to implement only the desired phases on the controlled PSs. We demonstrate now that this strategy does not provide an accurate control of the PIC. On a simulated 12-mode Clements interferometer, the generated relation (see Methods "Simulation benchmark of the training process") featuring a rectangular extended crosstalk matrix $\CTWOEXT$ is inverted using the pseudo-inverse in our solver. The calculated voltage solution yields on average $\SI{6}{mrad}$ errors on controlled phases and $\SI{140}{mrad}$ residual phase shifts on induced PSs. As a result, the amplitude fidelity (see Eq.~\ref{eq:fidelity}) of implemented matrices with respect to the target is degraded to $\FA=\SI{97.7(6)}{\%}$ (error bar is one standard deviation) as shown on Fig.~\ref{fig:reduction}d (blue histogram). In contrast, the case of simulated PICs without induced PSs, i.e.\ with square crosstalk matrices, our solver achieves $\FA=\SI{100}{\%}$. Mitigation of induced PSs thus requires to transform the rectangular crosstalk matrix $\CTWOEXT$ into an invertible square matrix.

To mitigate crosstalk in the extended framework, we introduce in Subsection \ref{subsec:matrix_transformations} a method called \textit{crosstalk matrix reduction}, which turns rectangular crosstalk matrices into square invertible crosstalk matrices. The matrix reduction process uses invariant phase transformations. We lay the foundations for finding such phase transformations in Subsection \ref{subsec:circuit_rewriting} using circuit rewriting rules. From the rewriting rules, we devise an efficient algorithm for finding invariant phase transformations on a given circuit (see App.~\ref{app:phase_algo}). Subsection \ref{subsec:criteria} gives a practical criterion for straightforwardly certifying the ability of a given PIC to fully cancel crosstalk.

\subsection{Matrix reduction from invariant phase transformations mitigates induced PSs}
\label{subsec:matrix_transformations}

\begin{figure*}[ht]
    \centering
    \includegraphics[width=\linewidth]{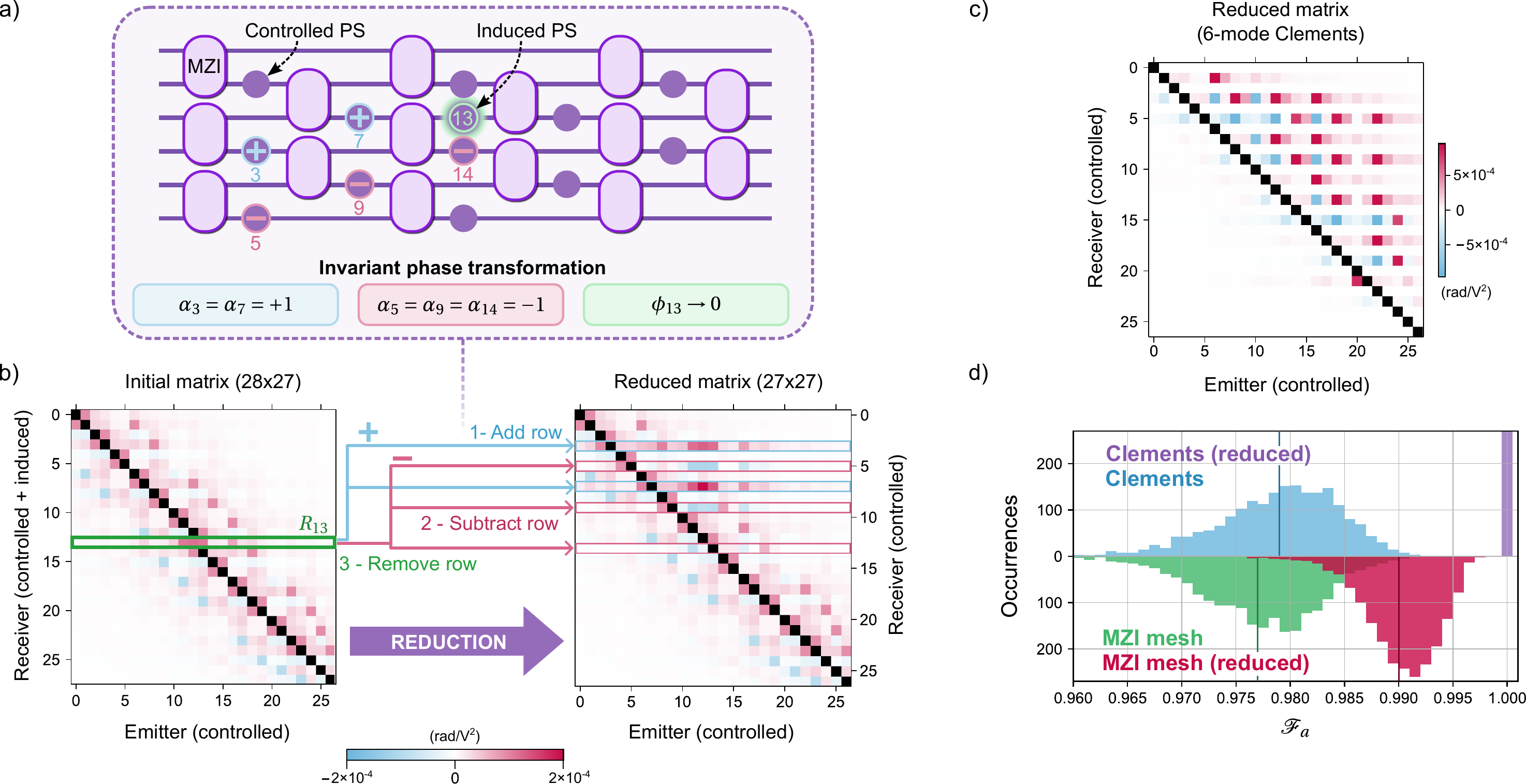}
    \caption{\textbf{Mitigation of induced phase shifters (PSs) by matrix reduction.}
    All input and output ports are here assumed to be phase-invariant (see Section \ref{subsec:matrix_transformations}).
    \textbf{a) + b)} Matrix reduction is illustrated on a 6-mode Clements interferometer featuring a single induced PS. The circuit features $\NPS=28$ PSs, among which $\NCPS=27$ are controlled. We consider that all of the input and output ports of the circuit are phase-invariant (see Section \ref{subsec:matrix_transformations}) 
    \textbf{a)} Schematic of the circuit. The purple squares represent Mach-Zehnder interferometers (MZIs, each featuring one controlled PS), and the purple disks depict controlled PSs. The induced PS (green disk) indexed as PS 13 implements a phase shift $\phi_{13}$. The circuit admits an invariant phase transformation of the form Eq.~\ref{eq:pit} which consists in setting $\phi_{13}$ to 0, adding $\phi_{13}$ to the phase of PSs indexed 3 and 7 (PSs with a "+" sign) and subtracting $\phi_{13}$ from the phase of PSs indexed 5, 9 and 14 (PSs with a "-" sign). The invariant phase transformation is of the form Eq.~\ref{eq:pit} with coefficients $\alpha_3=\alpha_7=+1$ and $\alpha_5=\alpha_9=\alpha_{14}=-1$ for $\phi_{13}\rightarrow 0$.  
    \textbf{b)} Illustration of the reduction process. By Theorem \ref{thm:reduction}, the invariant phase transformation shown in a) is associated to an invariant crosstalk matrix transformation. We select the 13-th row (green frame), denoted $R_{13}$ of the initial rectangular crosstalk matrix of size $\NPS\times\NCPS$. The dominant diagonal coefficients with approximate value $\SI{0.034}{rad\per V^2}$ are masked for readability. According to the coefficients $\alpha_i$, $R_{13}$ is added ($\alpha_i=1$, blue frames) to or subtracted ($\alpha_i=-1$, red frames) from the rows of the initial matrix. The process yields a square reduced matrix of size $\NCPS\times\NCPS$ after deletion of $R_{13}$.
    \textbf{c)} Reduced crosstalk matrix of the 6-mode Clements interferometer to reduce the rectangular crosstalk matrix displayed in Fig.~\ref{fig:model_extension}d. The crosstalk matrix row deletion procedure of Theorem \ref{thm:reduction} is applied to every induced PS.
    \textbf{d)} Histogram of amplitude fidelities with vertical bars indicating the average value. We simulate two 12-mode circuits featuring induced phase shifters, whose initial rectangular crosstalk matrix is generated from the circuit layout in the extended crosstalk framework (see Methods "Simulation benchmark of the training process"), i.e.\ every waveguide either has a controlled or an induced PS. For each circuit, the amplitude fidelity with respect to target matrices is measured on 2000 target phase configurations (see Methods "Amplitude fidelity measurement"). Upper half: Clements interferometer with rectangular crosstalk matrix (blue) and with reduced square matrix (purple, all values are 1 up to numerical errors). Lower half: mesh of MZIs (Clements interferometer whose external controlled PSs have been converted into induced PSs, as for Fig. \ref{fig:convergence}d) with rectangular crosstalk matrix before (green) and after (red) reduction. 55 of the initial 186 induced PS cannot not be removed from the MZI mesh, thus the reduced matrix of size $121\times66$ remains rectangular. 
    }
    \label{fig:reduction}
\end{figure*}

Crosstalk matrix reduction, which plays a pivotal role in induced PS mitigation, hinges on the existence of \textit{invariant PIC transformation}. Invariant PIC transformations are modifications of the PIC properties that leave the \textit{measurable outcomes} of the physical experiment invariant, while keeping the same voltages applied to the PIC. The measurable outcomes may be the light intensity, photon countrates or photon coincidence counts

Theorem \ref{thm:reduction} allows us to remove induced PSs from a given PIC by applying an invariant transformation on the PIC.

\bigskip

\hrule

\begin{theorem}
\label{thm:reduction} (Crosstalk matrix row deletion)

    \medskip

    Consider a PIC with PSs labeled $1, 2, ..., n$ (including controlled and induced PSs) and extended crosstalk matrix $\CTWOEXT$. We denote $\phi_i$ the phase shift value implemented by the i-th PS. 

    \medskip
    
    Suppose that the k-th PS is an induced PS and that there exists an invariant phase transformation of the form
    \begin{align}
        \label{eq:pit}
        \begin{split}
            \phi_i &\longrightarrow \phi_i + \alpha_i \phi_k \hspace{0.2cm} \text{for} \hspace{0.2cm} i\neq k \\
            \phi_k &\longrightarrow 0
        \end{split}
    \end{align}
    that holds for all values of $\phi_k\in[0, 2\pi[$ and where the coefficients $\alpha_1,...,\alpha_n$ are in $\mathds{R}$. 
    
    \medskip
    
    Then we may apply the following invariant transformation to the PIC:
    \begin{enumerate}
        \item Apply the transformation  
        \begin{equation}
            R_i \longrightarrow R_i + \alpha_i R_k \hspace{0.2cm} \text{for} \hspace{0.2cm} i\neq k
        \end{equation}
        to the crosstalk matrix $\CTWOEXT$ of the PIC, where $R_i$ designates the i-th row of $\CTWOEXT$.
        \item Delete $R_k$.
        \item Remove the k-th PS from the circuit.
    \end{enumerate}
\end{theorem}

\hrule

\bigskip

In Theorem \ref{thm:reduction}, \textit{invariant phase transformations} are modifications of the implemented PIC phase shifts that leave the measurable outcomes invariant. Invariant phase transformations are a property both of the PIC and the optical setup. For instance, if the PIC is followed by light intensity detectors, then applying additional phase shifts to the output modes of the PIC does not produce any measurable effect. We then say that the output ports are \textit{phase-invariant}. Similarly, if the PIC is used to manipulate single photons as in \cite{Maring2024}, then the input ports are phase-invariant as well. Note that induced and controlled PSs located on phase-invariant input and output ports may be discarded.

The crosstalk matrix row deletion procedure of Theorem \ref{thm:reduction} is illustrated on Fig.~\ref{fig:reduction}a,b. Note that every usage of Theorem \ref{thm:reduction} deletes a row from the crosstalk matrix and removes effectively an induced PS from the circuit. By applying the procedure to each induced PS of a given PIC, a rectangular extended crosstalk matrix $\CTWOEXT$ may be reduced into a square matrix, depending on the interferometer mesh. For instance, Fig.~\ref{fig:reduction}c shows that the extended rectangular crosstalk matrix of a 6-mode Clements interferometer \cite{Clements2016} reduces to a square matrix, assuming phase-invariant input and output ports. Fig.~\ref{fig:model_extension}d displays the initial rectangular crosstalk matrix. Note that the reduction process introduces artificial long-range crosstalk in the reduced matrix to compensate for the removal of induced phase shifters.

To perform the reduction, only invariant PIC transformations are applied to the PIC. As a result, the reduction process is an invariant PIC transformation as well. In other words, the initial rectangular and the reduced crosstalk matrices yield the same measurable outcomes for a given list of voltages $\vec{V}$.

If the reduced crosstalk matrix is square, it is typically also invertible (see Methods). An invertible reduced crosstalk matrix allows to exactly invert the phase-voltage relation and recover the full PIC control accuracy as demonstrated in the upper half of Fig.~\ref{fig:reduction}d.

However, the reduction process yields a square matrix, only if we can find an invariant phase transformation for each induced PS. We give in Subsection \ref{subsec:criteria} a criterion on the circuit for this condition to be fulfilled. For instance, consider a mesh consisting only of MZIs without controlled PSs between them (e.g.\ the circuit of Fig.~\ref{fig:reduction}a without purple dots). The initial rectangular matrix can then only be partially reduced, resulting in a reduced matrix that remains rectangular. Partial reduction nevertheless provides a control accuracy improvement over the initial matrix (see lower half of Fig.~\ref{fig:reduction}d).

\subsection{Finding invariant phase transformations from local circuit rewriting rules}
\label{subsec:circuit_rewriting}

\begin{figure*}[ht]
    \centering
    \includegraphics[width=\linewidth]{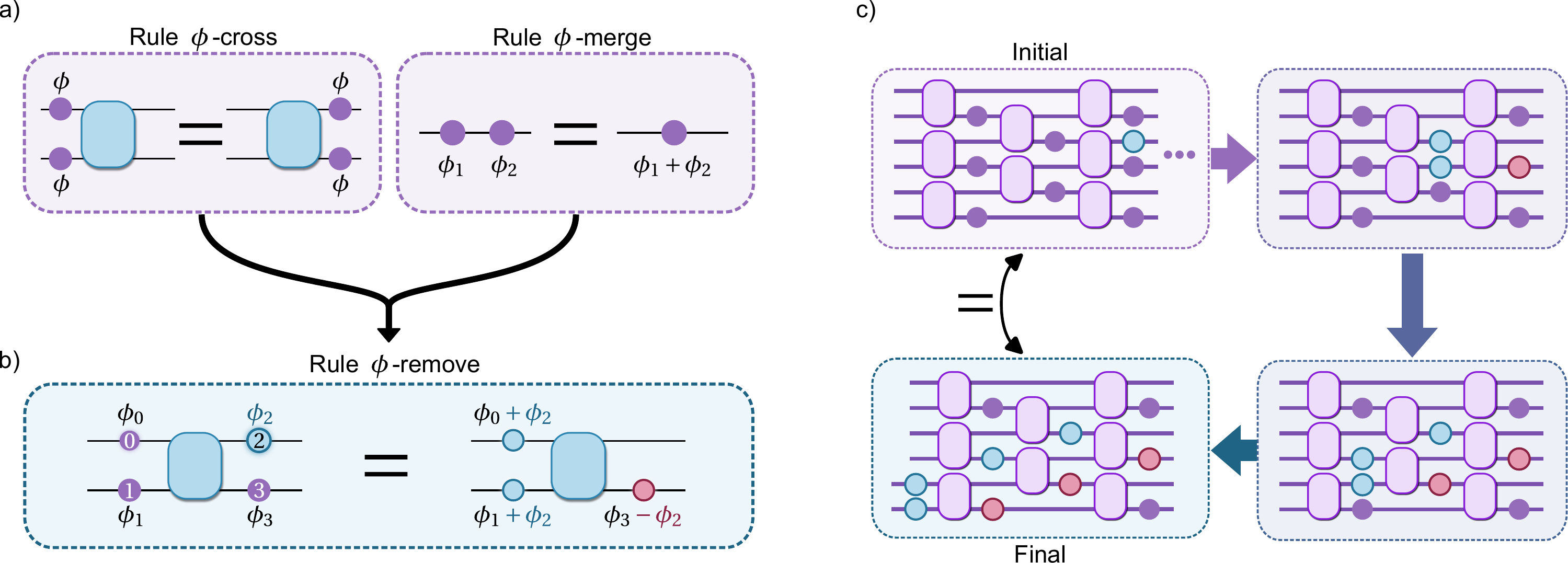}
    \caption{
    \textbf{Local circuit rewriting rules are used to find invariant phase transformations.} Disks represent phase shifters (PSs). Equality between two circuits means that both implement the same unitary matrix acting on the photons.
    \textbf{a)} Basis rules $\phi$-cross and $\phi$-merge from which the rule $\phi$-remove \textbf{b)} is derived. The blue rectangle represents any linear on-chip component on two modes such as a beamsplitter or a Mach-Zehnder interferometer (MZI). The phase shift implemented by each PS is displayed near its corresponding disk.
    \textbf{c)} Derivation of the invariant phase transformation of Fig.~\ref{fig:reduction}a. Purple rectangles are MZIs. Purple disks are controlled PSs. In the initial state, the blue disk is an induced PS with a phase $\psi$ to move out of the circuit. The three dots on the right indicate that the circuit has been truncated for readability. The induced phase shift is moved across the circuit by repeatedly applying the rule $\phi$-remove from b). The rule $\phi$-remove is applied to cross an MZI. In the intermediate and final states, blue (resp. red) disks represent PSs whose phase has been increased (resp. decreased) by $\psi$. The process ends when only phase shifts outside of the circuit remain. They are discarded in Fig.~\ref{fig:reduction}a because we assume phase-invariant input and output ports in this case (see Section~\ref{subsec:matrix_transformations}).
    }
    \label{fig:rewriting}
\end{figure*}

Matrix reduction relies on the existence of invariant phase transformations. Such invariant phase transformations can be found by moving the induced PSs out of the circuit using local circuit rewriting rules. In particular, we use the two basis rules $\phi$-cross and $\phi$-merge depicted on Fig.~\ref{fig:rewriting}a and also introduced in \cite{Bell2021, Pont2022, Clément2022, Heurtel2024}. 

From the rules $\phi$-cross and $\phi$-merge, we derive the rule $\phi$-remove which is used to move a phase shift by modifying only adjacent phase shifts. PSs are said to be \textit{adjacent} when they are in the configuration displayed in Fig.~\ref{fig:rewriting}b. 

The rule $\phi$-remove is an invariant phase transformation of the form Eq.~\ref{eq:pit}. With the notation of Fig.~\ref{fig:rewriting}b, the coefficients of the transformation are $\alpha_1 = -1$ and $\alpha_2=\alpha_3=1$ for $\phi_0\rightarrow0$. Thus, if the PS labeled "2" represents an induced PS and PSs labeled "0", "1" and "3" are controlled PSs, then the crosstalk matrix row deletion procedure (Theorem \ref{thm:reduction}) can be applied to remove the induced PS from the circuit and delete its associated row in the crosstalk matrix. If an induced PS is not adjacent to three controlled PSs, then the $\phi$-remove rule must be repeatedly applied until all created phase shifts have been merged with controlled PSs or moved out of the circuit (see Fig.~\ref{fig:rewriting}c). App.~\ref{app:phase_algo} introduces an efficient and generic phase simplification algorithm to find the invariant phase transformation associated to each induced PS, which are then used to reduce crosstalk matrices following Theorem \ref{thm:reduction}.

In the particular case of Reck interferometers \cite{Reck1994}, \cite{Clément2022} showed that there exists a set of deterministic rules , i.e.\ each step in the process has only one applicable rule ensuring an unambiguous path of execution. These deterministic rules can be used to remove all the induced PSs in Reck interferometers.

To find phase invariant transformations, we only consider linear relationships between PSs found via the $\phi$-remove rule (see Fig.\ \ref{fig:rewriting}b) It would be possible to remove additional induced PSs from circuits by also considering nonlinear relationships between phases (e.g.\ axiom E2 in \cite{Heurtel2024}). This would however greatly complexify the transformations applied to the crosstalk matrix, ultimately rendering the phase-voltage equation challenging to solve.

\subsection{Criterion for crosstalk-robust interferometers: acyclic pruned graphs}
\label{subsec:criteria}

\begin{figure*}[ht]
    \centering
    \includegraphics[width=\linewidth]{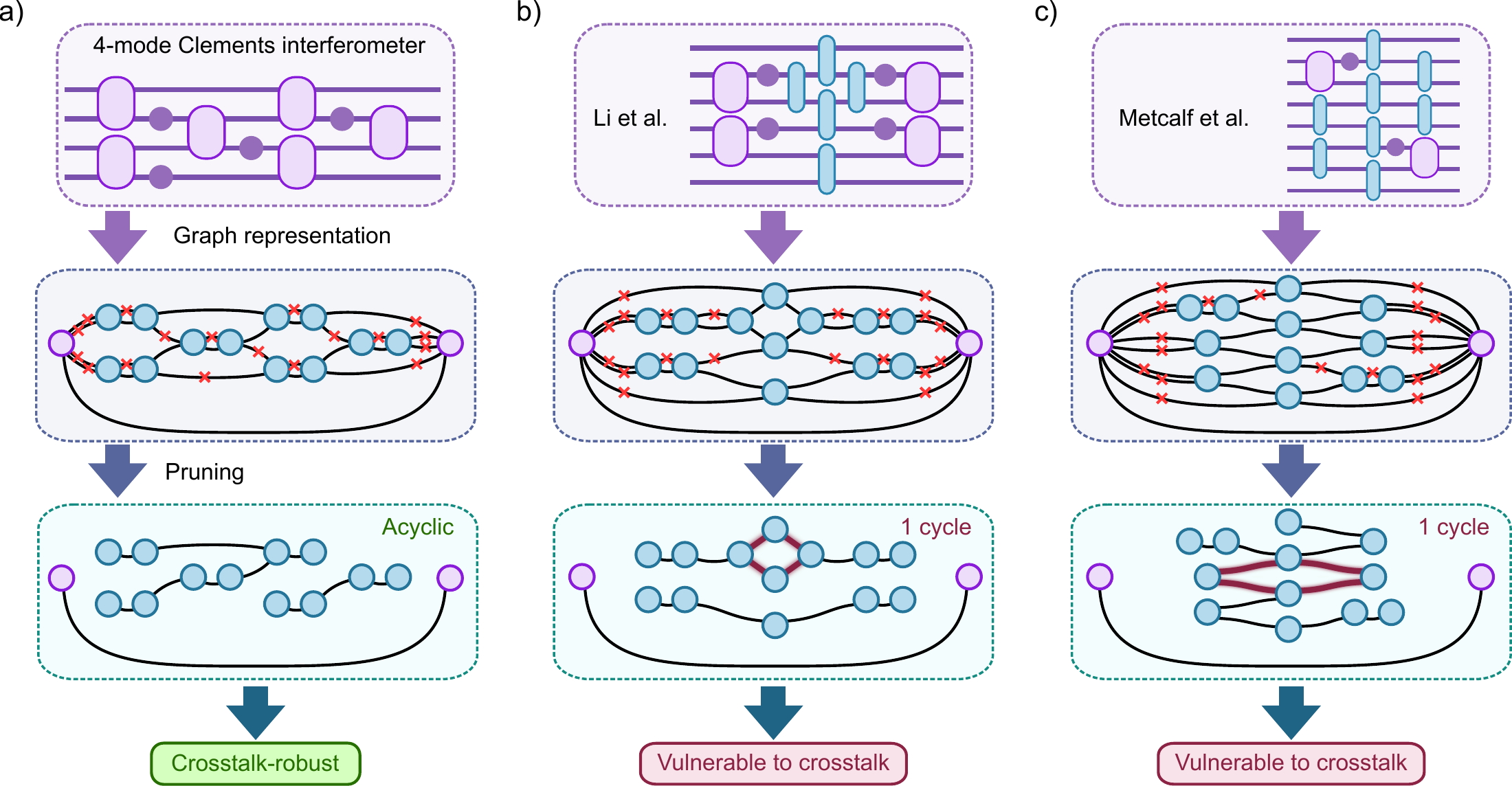}
    \caption{
    \textbf{Certification of robustness to crosstalk using graph theory}
    The crosstalk robustness criterion (Theorem \ref{thm:robust}) is illustrated for \textbf{a)} a 4-mode Clements interferometer \cite{Clements2016} and the interferometers of \textbf{b)} \cite{Li2013} and \textbf{c)} \cite{Metcalf2014}. Each circuit is drawn in the respective top panel, with controlled phase shifters (PSs) represented as purple disks, beamsplitters as blue rectangles and Mach-Zehnder interferometers with the PS on the top mode as purple rectangles. Each circuit is associated to a mathematical graph, where the blue nodes symbolize the circuit beamsplitters, the purple nodes are the input and output nodes and the graph edges are laid out to reproduce the circuit waveguide mesh. Here we assume that all the ports are phase invariant (see Section \ref{subsec:matrix_transformations}). Phase invariant input and output ports and waveguides featuring a controlled PS are marked with a cross. The graph is then pruned according to Section \ref{subsec:criteria} by deleting all the marked edges. The pruned graph of a) is acyclic which proves that the 4-mode Clements interferometer is crosstalk-robust. On the contrary, the pruned graphs of b) and c) contain one cycle, thus the interferometers cannot entirely cancel crosstalk.
    }
    \label{fig:cycles}
\end{figure*}

We establish in this subsection a graphical criterion for certifying the robustness of a PIC to crosstalk. PICs that are not crosstalk-robust are not suited as experimental platforms, as they do not have the ability to suppress crosstalk entirely. A PIC is \textit{crosstalk-robust} when its initial rectangular crosstalk matrix can be fully reduced to a square matrix following Section \ref{subsec:matrix_transformations}, using linear invariant phase transformations derived from the circuit rewriting rule $\phi$-remove from Section \ref{subsec:circuit_rewriting}. 

The criterion does not rely on the phase simplification algorithm of App.~\ref{app:phase_algo}, and provides intuition on why certain meshes fail to be crosstalk-robust. We demonstrate how to apply our graphical criterion in Fig.~\ref{fig:cycles}. The PIC to investigate is represented as a mathematical graph. The nodes represent the on-chip beamsplitters and the edges interconnecting the nodes symbolize the PIC waveguides. The edges corresponding to PIC waveguide section featuring a controlled PS are marked with a cross. The graph features two additional nodes, called \textit{input} and \textit{output} nodes which are connected by an edge. Every input (resp. output) port of the circuit is connected to the input (resp. output) node. Each of these edges is marked with a cross if the port is phase invariant. We obtain a \textit{pruned graph} by deleting every edge marked with a cross. The pruned graph encodes the robustness of the PIC to crosstalk as stated in Theorem \ref{thm:robust}, proven in App.~\ref{proof:robust}. 

\bigskip
\hrule 

\begin{theorem}[Crosstalk-robustness criterion]
\label{thm:robust} 
A PIC is crosstalk-robust if and only if its associated pruned graph is acyclic.
\end{theorem}

\hrule 
\bigskip

The criterion is illustrated in Fig.~\ref{fig:cycles} on three examples with phase-invariant input and output port. App.~\ref{app:phase_dependent} provides an example where phase-invariance is not assumed. Following Theorem \ref{thm:robust}, the Reck \cite{Reck1994}, Clements \cite{Clements2016} and Bell-Walmsley \cite{Bell2021} universal interferometers are crosstalk-robust. On the contrary, specialized interferometers such as \cite{Shadbolt2012, Li2013, Metcalf2014, Peruzzo2014} are not crosstalk-robust for instance. Remarkably, we can highlight that universal schemes are not necessarily crosstalk-robust. Indeed, an MZI followed by two beamsplitters is universal on two modes, but not crosstalk-robust.

A pruned graph containing cycles can be made acyclic by removing additional edges. The  minimal number of edges that must be removed corresponds to the circuit rank, as defined in graph theory \cite{berge2001}. In other words, a non-crosstalk-robust interferometer can be made resilient by adding $r$ controlled PSs to the circuit, where $r$ is the circuit rank of its pruned graph. The placement of these additional PSs is discussed in App.~\ref{proof:maximal}.

Robustness to crosstalk also sheds light on the results of Fig.~\ref{fig:convergence}c,d. In Fig.~\ref{fig:convergence}c, the restricted MLM equipped with a square crosstalk matrix is able to faithfully learn the behavior of the Clements interferometer. This is because Clements interferometers are crosstalk-robust. Thus the crosstalk dynamics of Clements interferometers can be encapsulated into a square crosstalk matrix, which the restricted MLM can then directly learn. On the contrary in Fig.~\ref{fig:convergence}d, the restricted MLM cannot converge in the case of a generic mesh of MZIs which is not crosstalk-robust, because its crosstalk is necessarily described by a rectangular matrix.

\section{Experimental validation}
\label{sec:validation}

\begin{figure*}[ht]
    \centering
    \includegraphics[width=0.95\linewidth]{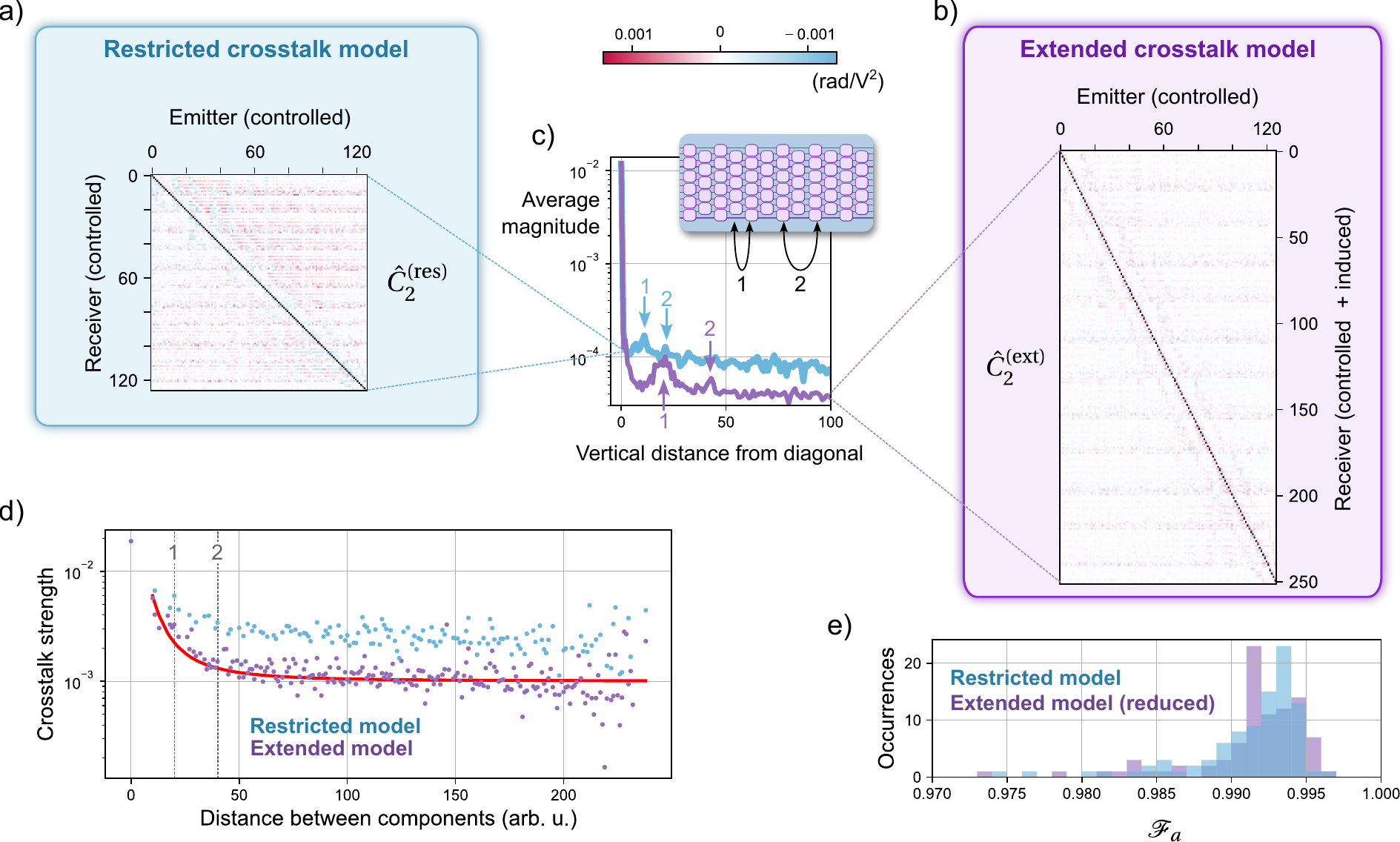}
    \caption{
    \textbf{Experimental characterization of a 12-mode Clements interferometer with restricted and extended crosstalk machine learning models (MLMs).}
    \textbf{a)} (resp. \textbf{b)}) Estimated crosstalk matrix $\CTWORESHAT$ (resp. $\CTWOEXTHAT$) of the trained MLM equipped with a restricted (resp. extended) crosstalk model. The self-heating coefficients on the diagonal are masked.
    \textbf{c)} Averaged magnitude of matrix elements as a function of vertical distance from the diagonal along vertical sections of $\CTWORESHAT$ and $\CTWOEXTHAT$. Arrows labeled "1" and "2" indicate distances from the diagonal corresponding to unit cell column jumps as indicated in the inset.
    \textbf{d)} Crosstalk strength in $\CTWORESHAT$ and $\CTWOEXTHAT$ as a function of the distance between components. Crosstalk strength is defined as the crosstalk coefficient value divided by the average diagonal coefficient. The distance is estimated from a 2D layout of the physical device described (see Methods "Simulation benchmark of the training process"). The red continuous curve shows the crosstalk-versus-distance function used for the simulations, with an additional offset of 1e-3 for clarity due to the log scale. The vertical gray lines labeled "1" and "2" indicate the distance between consecutive unit cell columns (see inset of c)).
    \textbf{e)} The PIC control accuracy provided by the restricted and extended MLM characterizations is measured for 100 phase configurations (see Methods "Amplitude fidelity measurement"). The extended crosstalk matrix $\CTWOEXTHAT$ is reduced to a square matrix prior to the measurement following Section \ref{sec:mitigation} to enable crosstalk mitigation. The accuracy is measured by the amplitude fidelity $\FA$ between with measured and expected output light intensity distributions. The average value for the restricted (resp. extended) MLM is $\SI{98.7(2)}{\%}$ (resp. $\SI{98.6(2)}{\%}$) (error bar is the statistical uncertainty on the 100 experiments).
    }
    \label{fig:experiment}
\end{figure*}

We proceed with the comparison of restricted and extended MLMs with a physical PIC. We characterize and control a 12-mode Clements interferometer featuring 126 reconfigurable phase-shifters and 132 beamsplitters \cite{Taballione2021}. As described in Section \ref{sec:charac}, we acquire $\approx$ 16 000 training samples and 4 000 test samples on the PIC, corresponding to a data-to-parameter ratio of 1 for the restricted crosstalk model and a train/test dataset ratio of 80/20.

The PIC is first characterized with a restricted MLM following the procedure of \cite{Fyrillas2024b}, consisting in the stages:
\begin{itemize}
    \item V-IFM (voltage interference fringe measurement): the estimated passive phases $\hvec{c}_0$ and self-heating coefficients (diagonal elements of the estimated crosstalk matrix $\CTWORESHAT$) are coarsely approximated by sweeping the voltage of each controlled PS and fitting the resulting interference fringes.
    \item ML (machine learning stage): the estimated crosstalk matrix $\CTWORESHAT$, beamsplitter reflectivity values $\hvec{R}$ and output transmissions $\hvec{T}_\text{out}$ are updated via a gradient descent algorithm. Similarly to Section \ref{sec:charac}, the algorithm seeks to minimize the mean squared error between the measured output light intensity distributions $\vec{p}$ in the training dataset and the distributions $\hvec{p}$ predicted by the model.
    \item $\phi$-IFM (phase interference fringe measurement): the estimated passive phases $\hvec{c}_0$ are updated by sweeping successively each PS phase from 0 to $2\pi$. This measurement makes use of the knowledge acquired in the preceding ML stage by the MLM to 1) compensate crosstalk for improved phase shifter control using the estimated crosstalk matrix $\CTWORESHAT$ and 2) to generate the expected interference fringes using the learned beamsplitter reflectivity values $\hvec{R}$ and output losses $\hvec{T}_\text{out}$. The measured fringes are compared to the expected ones to deduce the passive phase update. 
    \item The protocol iterates between ML and $\phi$-IFM stages until the prediction accuracy of the model stagnates, measured by the average TVD (see Section \ref{sec:charac}) on the test dataset.
\end{itemize}
The restricted MLM achieves $\text{TVD}_\text{test}^\text{(res)} = \SI{7.5}{\%}$ at the end of the protocol, after 2 (ML + $\phi$-IFM) iterations. The estimated crosstalk matrix $\CTWORESHAT$ is displayed in Fig. \ref{fig:experiment}a. The difference between our value for $\text{TVD}_\text{test}$ and the one of \cite{Fyrillas2024b} ($\text{TVD}_\text{test}^\text{(res)} = \SI{2.9}{\%}$) resides in our use of an attenuated laser instead of a true single-photon source as the light source (see Methods).

We then equip the MLM with an extended phase-voltage relation (see Eq.~\ref{eq:phase_voltage_extended}) and repeat the characterization process, with 2 (ML + $\phi$-IFM) iterations as well. For this, we introduce a modification in the protocol where the rectangular crosstalk matrix $\CTWOEXTHAT$ is reduced to a square crosstalk matrix before entering each $\phi$-IFM stage following our crosstalk matrix reduction procedure from Section \ref{sec:mitigation}. The reduction is necessary to effectively mitigate crosstalk when sweeping each PS in phase. The extended crosstalk model achieves $\text{TVD}_\text{test}^\text{(ext)} = \SI{7.6}{\%}$. As predicted from our simulations (see Fig \ref{fig:convergence}c) and justified in Section \ref{subsec:criteria}, the restricted and extended MLMs yield similar $\text{TVD}_\text{test}$ values for Clements interferometers. The estimated extended crosstalk matrix $\CTWOEXTHAT$ is shown in Fig.~\ref{fig:experiment}b.

We quantify the PIC control accuracy by the amplitude fidelity $\FA$ between measured and implemented phase configurations on the PIC. Empirically, we observe that $\FA \approx 1 - \text{TVD}_\text{test}^2$. Consequently, the comparable $\text{TVD}_\text{test}$ achieved by the restricted and extended MLMs implies that using the estimated physical parameters of either model yields a similar PIC control accuracy. Indeed, the amplitude fidelity $\FA$ on the physical PIC is the same for both models within error bars (see Fig.~\ref{fig:experiment}e). 

The total time spent in the 2 ML stages is 3.9 minutes for the restricted MLM (see Methods "Simulation benchmark of the training process" for the duration of an epoch). On the other hand, the extended MLM offers a speed up as observed in Fig.~\ref{fig:convergence}b, requiring only 1.5 minutes to reach the same $\text{TVD}_\text{test}$. 

The restricted MLM is able to account for induced phase shifters in the case of Clements interferometers, as simulated in Fig.~\ref{fig:convergence}b and discussed in Section \ref{subsec:criteria}. The restricted MLM achieves this by introducing long-range crosstalk in the estimated restricted matrix $\CTWORESHAT$ (see Fig.~\ref{fig:experiment}a). The resulting estimated matrix is akin to the reduced version of an extended crosstalk matrix (see Fig.~\ref{fig:reduction}a). Due to the artificial long-range crosstalk, the average off-diagonal amplitude of the estimated restricted crosstalk matrix $\CTWORESHAT$ is $\approx \SI{1e-4}{}$, whereas it is only $\approx \SI{5e-5}{}$ for the extended crosstalk matrix $\CTWOEXTHAT$. As a consequence, physical short-range crosstalk is obscured in the restricted crosstalk matrix $\CTWORESHAT$, whereas it is clearly visible in the extended matrix $\CTWOEXTHAT$. In particular, the extended matrix shows the presence of crosstalk between adjacent MZI columns (see Fig.~\ref{fig:experiment}c). Furthermore, reconstructing crosstalk as a function of distance between components in Fig.~\ref{fig:experiment}d shows that the crosstalk is more short-ranged in the extended matrix $\CTWOEXTHAT$ than in the restricted matrix $\CTWORESHAT$. Thus, extended crosstalk models capture a more realistic picture of crosstalk, offering a valuable method for assessing PIC design and fabrication improvements in large-scale integrated structures.

\section{Conclusion and discussion}

We propose an extended crosstalk model including induced PSs, providing a systematic and physically faithful approach to modeling PS crosstalk in PICs. We use machine learning to estimate crosstalk matrices of physical devices and demonstrate in simulations that MLMs equipped with an extended crosstalk model generally converge faster than MLMs equipped with a restricted model. This behavior is demonstrated on Clements interferometers \cite{Clements2016} , and further confirmed in separate simulations on Bell-Walmsley interferometers \cite{Bell2021} (see App.~\ref{app:bell_comparison}). The restricted and extended model require the same number of data samples for training. Additionally, the extended MLM successfully converges on a larger class of interferometers than the restricted MLM. 

Our crosstalk mitigation framework is able to achieve optimal control of PICs with known universal-scheme interferometers, such as the Reck \cite{Reck1994}, Clements \cite{Clements2016}, and Bell-Walmsley \cite{Bell2021} schemes, despite the presence of induced PSs. However, scalability beyond 50 modes, or 25 dual-rail path-encoded qubits, with universal interferometers remains a challenge due to the large number of on-chip components required. Consequently, quantum information processing with a larger number of photonic qubits will likely necessitate specialized interferometers. Our mitigation framework reveals a significant caveat in this approach: some PIC interferometer designs are fundamentally flawed, preventing effective crosstalk cancellation. 

To enforce robustness to crosstalk, it is possible to equip every waveguide of a PIC with a controlled PS at the design stage. Such an approach to PIC design would however affect negatively the footprint, control complexity, power consumption, heat dissipation capabilities and optical transmission of integrated devices. We establish an accessible criterion that certifies the robustness of a given PIC to crosstalk. As such, our criterion can be used to design PICs with reduced numbers of controlled PSs, while maintaining a sufficient number of degrees of freedom to mitigate crosstalk.

Experimentally, the extended MLM proves effective in capturing a physically faithful representation of crosstalk within physical devices. This capability is particularly valuable for PIC design and fabrication processes, offering a direct benchmark for crosstalk reduction strategies in large-scale PICs, in which accessing individual physical parameters is challenging.

Our extended crosstalk model, conclusions on crosstalk characterization and mitigation, as well as the layout rules for designing interferometers resilient to crosstalk are valid for any PS technology.

\section*{Methods} 

\subsection*{Thermal crosstalk simulations}

We use the HEAT and Finite Difference Eigenmode (FDE) solvers of the Ansys Lumerical suite to perform thermal and electromagnetic simulations. We consider the circuit cross section described in Fig.~\ref{fig:cross_section}b. Further information about the parameters and the results of the simulation can be found in the App.~\ref{app:th_sim}.

The thermal crosstalk arising when a controlled PS is actuated is estimated by retrieving the temperature profile in the circuit. The steady state temperature reached in each point of the cross section can be calculated by numerically solving the time-independent heat equation
\begin{equation}
    -k(x,y)\cdot\nabla^2T(x,y)=Q(x,y),
    \label{eq: heat}
\end{equation}
where $k$ [W/(m $\cdot$ K)] is the thermal conductivity of the material, $\nabla^2T$ is the Laplacian of the temperature $T$ [K], and $Q=\rho \cdot J^2$ [W/m\textsuperscript{3}] is the power per unit volume dissipated by Joule effect in a PS of resistivity $\rho$ and biased with a current density $J$.

Once the temperature has been computed, we find the new refractive index profile $n(x,y,T)$ in the cross section by using the thermo-optic coefficient $\partial n / \partial T$, which depends on the considered material - in this case silicon nitride (SiN) for the core and SiO\textsubscript{2} for the cladding:
\begin{equation}
    n(x,y,T)=n(x,y,T_0)+\frac{\partial n}{\partial T}\Big|_{(x,y)} dT(x,y).
    \label{eq: thermo-optic}
\end{equation}

The information about the index profile allows to retrieve the effective refractive index $n_\text{eff}$ of the fundamental TE (transverse electric) mode by using the FDE solver, which calculates the waveguide eigenmodes by solving Maxwell's equations in the waveguide cross section. Once $n_\text{eff}$ is known, the phase shift is eventually calculated for each waveguide as
\begin{equation}
    \Delta \phi = \phi(T)-\phi(T_0)=\frac{2\pi}{\lambda}\big[n_\text{eff}(T)-n_\text{eff}(T_0)\big]\; L\;,
    \label{eq: eff_index}
\end{equation}
with $T_0$ the temperature in passive conditions, and $L$ [mm] the length of the actuated controlled PS.

\subsection*{Simulation benchmark of the training process}
\label{met:sim_benchmark}

The hardware device is replaced by a simulated PIC with perfectly symmetric beamsplitters and unity input and output transmissions. The passive phases of the simulated PIC are all set to zero, which corresponds to an experimental situation where the passive phases have been precisely estimated. The distance $d$ between the components (MZIs and PSs) is algorithmically computed by drawing the interferometer mesh with a spacing of 10 units between waveguides and 10 units between consecutive components. A distance equal to 0 corresponds to the distance between the two PSs inside an MZI. The off-diagonal crosstalk coefficients are then given by the function
\begin{equation}
    f(d)=
    \begin{cases}
    0.5 / d^2 \hspace{0.4cm} \text{for} \hspace{0.4cm} d \neq 0 \\
    0.02  \hspace{0.75cm} \text{for} \hspace{0.4cm} d = 0
    \end{cases}
\end{equation}
which reproduces the experimentally observed crosstalk-versus-distance relationship (see Fig. \ref{fig:experiment}d). The diagonal coefficients are all taken to be equal to $0.034$. 

The MLMs are coded in \texttt{Python} using the \texttt{PyTorch} package. The \texttt{Adam} optimizer is used to minimize the mean square error of the model predictions on the training dataset with $\beta$ parameters (0.99, 0.9999) with learning rates $10^{-5}$ for a number of interferometer modes smaller than or equal to 14, $5\times10^{-6}$ for 16 modes and $1\times10^{-6}$ for 18 modes. The training/test dataset ratio is 80/20. We use an i7-12700 CPU with an NVIDIA T1000 8 GB GPU. For a 12-mode Clements interferometer, the restricted (resp.\ extended MLM) takes 235 ms (resp. 302 ms). The epoch times are significantly shorter than \cite{Fyrillas2024b} thanks to the GPU acceleration of matrix operations.

\subsection*{Experimental setup}

We use a 12-mode Clements interferometer with 126 controlled thermo-optic phase shifters and 132 beamsplitters, a continuous-wave laser at 925 nm and single-photon detectors. The PIC reconfiguration time is 2 s. The coherent light is strongly attenuated before the PIC by decoupling fibers such that the total countrate on all 12 detectors lies between $\SI{500}{kcounts\per\s}$ and $\SI{1.5}{Mcounts\per\s}$ to avoid detector saturation and validity of the single-photon regime. For every performed countrate measurement, the integration time is 1 s.

\subsection*{Amplitude fidelity measurement}

We describe our protocol for evaluating the circuit control accuracy via the amplitude fidelity, both for simulated and hardware experiments. A specified number of target phase configurations are applied successively on the physical or simulated PIC. Each phase shift is uniformly chosen between $0$ and $2\pi$. Note that the resulting unitary matrices are not uniformly distributed for the Haar measure. Each phase configuration is converted into voltages using a specified phase-voltage relation and our solver described in App.~\ref{app:solver}. For each phase configuration, the amplitude matrix implemented by the interferometer is measured by acquiring the output light intensity distribution for light injected successively into every other input port (light is injected into a single input port at a time). The component-wise square root of the measured distributions are the columns of the measured matrices. The expected matrices are computed from the phase configurations. 

For Fig.~\ref{fig:cross_section}, 2000 phase configurations are generated. The phase-voltage relation is solved taking into account only crosstalk between controlled PSs. The behavior of crosstalk in the simulated circuit is set to match the results from Fig.~\ref{fig:cross_section}c. Note that our crosstalk simulations were performed only on transverse cross-sections such as Fig.~\ref{fig:cross_section}b, thus we consider here that crosstalk only propagates vertically in the PIC drawn in Fig.~\ref{fig:cross_section}a. In reality, heat is expected to also diffuse horizontally. For each phase configuration, the amplitude fidelity $\FA$ (see Eq.~\ref{eq:fidelity}) of the implemented matrix is measured with respect to the target matrix computed from the phase configuration.

In the case of the experimental run, the estimated beamsplitter reflectivity values $\hvec{R}$ and the estimated output transmissions $\hvec{T}_\text{out}$ by the MLM are used as well to generate the expected matrices. The amplitude fidelity values are computed between the measured and expected matrices from Eq.~\ref{eq:fidelity}.

\subsection*{Reduced matrix invertibility}

If the reduced crosstalk matrix is square, it is typically invertible, although this depends on the number of modes and the strength of crosstalk. For small interferometers or weak crosstalk, the reduced matrix is diagonally dominant and therefore invertible. As the number of modes increases, the off-diagonal coefficients of the crosstalk matrix, which account for induced phase shifts, increase in magnitude. This increase can compromise the diagonal dominance of the reduced matrix.

We generate simulated crosstalk matrices using the process described in the Methods section ("Simulation benchmark of the training process"). For Clements interferometers with realistic crosstalk values, the reduced matrix is diagonally dominant up to 150 modes. We verified that up to 160 modes, the reduced matrix remains invertible, even though it is not diagonally dominant. In contrast, for a 50-mode interferometer with crosstalk increased 30-fold beyond experimental values, the reduced matrix becomes non-invertible.

\section*{Acknowledgements}

This work has been co-funded by the European Commission as part of the EIC accelerator program, under the grant
agreement 190188855. We acknowledge funding from the Plan France 2030 through the project ANR-22-PETQ-0013.

\bigskip

\section*{Competing Interests}

The authors declare that there are no competing interests.

\section*{Author contributions}

A.F. and J.S. formulated the original idea. A.F. conducted machine-learned characterization and carried out experimental validations. S.P. performed numerical crosstalk simulations. A.F. and N.H. designed mitigation strategies, with N.H. developing the algorithm and criteria for mitigation. A.F. and N.B. wrote the manuscript and designed the visualizations. N.B., N.M. and J.S. supervised the project.

\section*{Data and code availability}

The codebase and data generated as part of this work are available to research groups upon reasonable request to the corresponding author.

\bibliographystyle{naturemag}
\bibliography{biblio/biblio}

\newpage
\onecolumngrid
\appendix
\section{PARAMETERS AND RESULTS OF THE THERMAL SIMULATIONS}
\label{app:th_sim}

\begin{figure*}[ht]
    \centering
    \includegraphics[width=\linewidth]{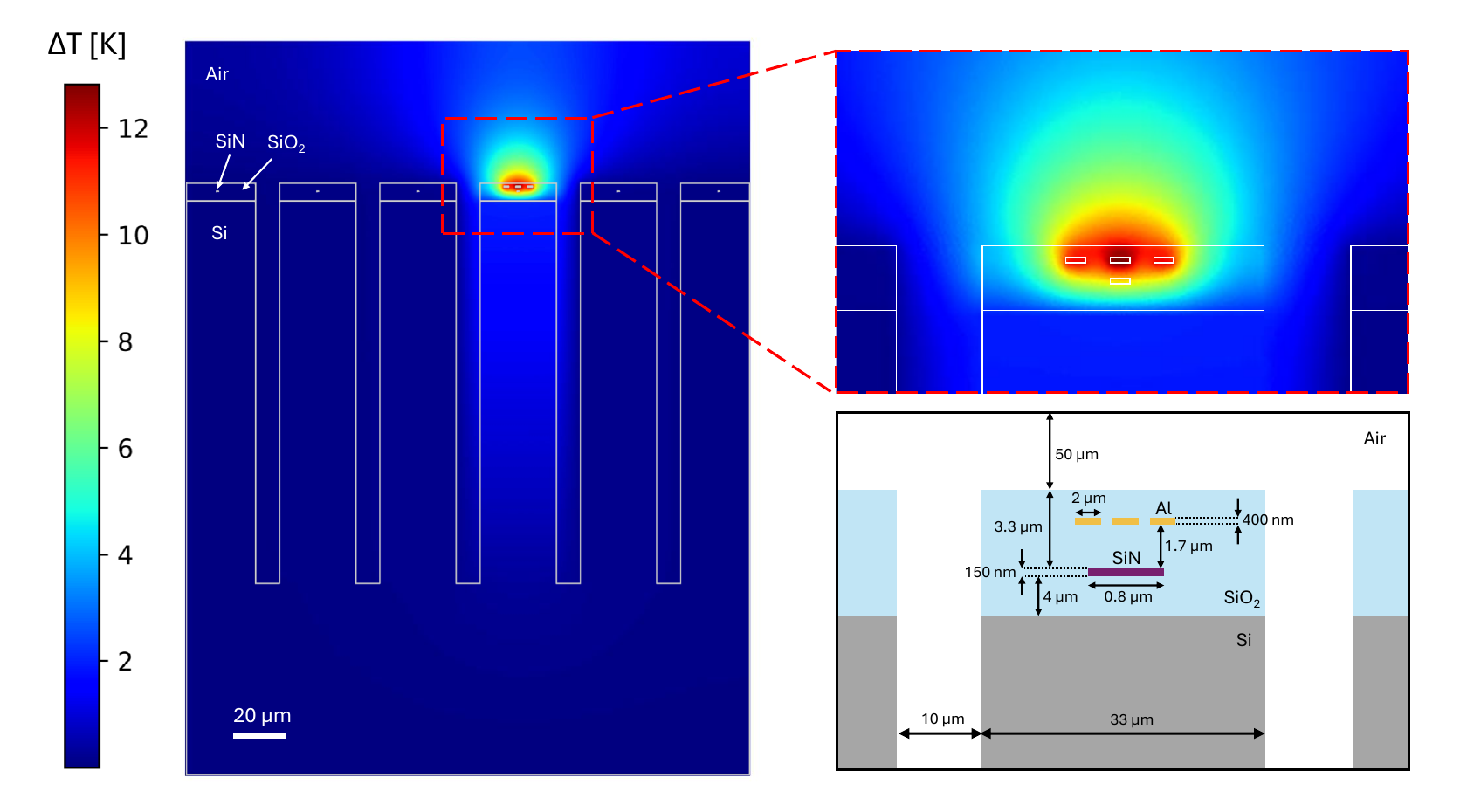}
    \caption{Simulated cross-section and temperature profile obtained when applying a bias current of 18 mA to the controlled phase shifter. In the insets, a detail of the temperature increase around the phase shifter, and a zoomed cross-section (not to scale) showing the main geometric parameters used for the simulation.}
    \label{fig:simulations}
\end{figure*}

A detail of the phase shifter (PS) cross section considered for the thermal and electromagnetic simulation is represented in the inset in Fig.~\ref{fig:simulations}. The SiN waveguide, with a thickness of 150 nm and a width of 800 nm, is placed under a PS composed of three aluminum alloy microheaters with a thickness of 400 nm, a width of 2 \textmu m and a pitch of 5 \textmu m. The distance between neighboring waveguides is 45 \textmu m. Insulating trenches, with a depth of 200 \textmu m, are etched in the silica (SiO$_2$) and silicon (Si) layers in the middle of each waveguides pair, and are considered filled with air. The total thickness of the device is set to 250 \textmu m. The simulation region also includes a layer of air on top of the oxide, to estimate the heat dissipation occurring in air by conduction; both convection and irradiation channels are neglected. 

A boundary condition of $T$ = 298 K is set at the bottom of the silicon layer, to simulate the behavior of an active temperature controller. Moreover, the size of the simulation area is chosen by performing convergence tests to ensure that the results are not affected by its choice. A non-uniform mesh is used in order to capture fast changes of temperature or electromagnetic fields only where necessary: for instance, the mesh size of the waveguides and the heaters is set to 20 nm, while for the silicon layer it is equal to 1 \textmu m. Convergence tests were performed as well to choose the adequate mesh size. The physical parameters used in the simulation, and in particular necessary for solving Eqs.~\ref{eq: heat}, \ref{eq: thermo-optic} and \ref{eq: eff_index}, are listed in Tab.~\ref{tab:materials}. The considered light wavelength is $\lambda$ = 925 nm. 

\begin{table}[ht]
    \centering
    \begin{tabular}{>{\centering\arraybackslash}p{2cm} wc{2cm} wc{2cm} wc{2cm} wc{2cm}}
    \hline\\
    Material & $n$ & $k$ [W/(m $\cdot$ K)] & $\rho $ [$\Omega$ $\cdot$ m] & $\partial n / \partial T$ [K\textsuperscript{-1}]\\[2pt]
    \hline\\
    SiN & 1.990 & 30 & - & 3.0 $\cdot$ 10\textsuperscript{-5}\\
    SiO\textsubscript{2} & 1.452 & 1.4 & - & 8.7 $\cdot$ 10\textsuperscript{-6}\\
    Al alloy & 1.771 + 8.641$i$ & 200 & 3.6 $\cdot$ 10\textsuperscript{-8} & -\\
    Si & 3.605 + 0.001$i$ & 130 & - & 1.8 $\cdot$ 10\textsuperscript{-4}\\
    Air & 1.000 & 0.03 & - & 1.0 $\cdot$ 10\textsuperscript{-6} \\[5pt]
    \end{tabular}
    \caption{Refractive index $n$, thermal conductivity $k$, electrical resistivity $\rho$ and thermo-optic coefficient $\partial n / \partial T$ at $T$ = 298 K and $\lambda$ = 925 nm for the materials considered in the simulation \cite{palik1998handbook, philipp1973optical}. If a value is not present, it is either not defined or not needed for the simulation.}
    \label{tab:materials}
\end{table}

In Fig.~\ref{fig:simulations}, we also show the simulated temperature profile when applying a bias current of 18 mA, corresponding to a voltage drop of 14 V on the 5 mm-long microheaters, which ensures a phase shift of $2\pi$ on the controlled waveguide. The maximum temperature difference induced in the material is equal to 12.5 K, with a temperature increase in the waveguide of 9 K. The simulation shows the insulating effect of the trenches, which are confining the heat in the column under the PS. It is also possible to notice how the remaining silicon layer is responsible for the heat transfer to the adjacent waveguides, due to the fact that the trenches are not etched through the whole thickness of the device.

\section{PHASE-VOLTAGE EQUATION SOLVER}
\label{app:solver}

Consider a photonic integrated circuit (PIC) with $\NCPS$ controlled phase shifters (PSs) and $\NPS$ PSs (controlled and induced). The general phase-voltage relation relating the implemented phase shifts $\vec{\phi}$ to the applied voltages $\vec{V}$ is of the form \cite{Fyrillas2024b}:

\begin{equation}
    \vec{\phi} = \sum_{k \geq 1} C_k \cdot \vec{V}^{\odot k} + \vec{c}_0,
\end{equation}
where $C_k$ are crosstalk matrices of size $\NPS\times \NCPS$, $^{\odot}$ is element-wise exponentiation and $\vec{c}_0$ is the vector containing the $\NPS$ passive phases. Additional features and constraints of the phase-voltage relation are
\begin{itemize}
    \item the components of $\vec{\phi}$ are defined up to $2\pi$
    \item the voltages assume positive values below a safe voltage operation bound.
    \item the diagonal of the crosstalk matrices dominates the off-diagonal terms
    \item for usual thermo-optic PSs, the order 2 term is dominant. Order 4, 6... terms may be added to account for heater resistance change with increasing temperature
\end{itemize}
Given $\vec{\phi} = \vec{\phi}\left(\vec{V}\right)$, we usually seek to invert the relation to obtain a set of voltages $\vec{V}$ that implements a target phase configuration $\vec{\phi}_\text{target}$. General nonlinear matrix equations are notoriously hard to solve, nevertheless our custom solver harnesses the characteristic traits of the phase-voltage relation to provide an approximate solution. 

We use an iteration-based solver. At each iteration, the voltage solution is updated according to the following rule inspired from perturbation theory:

\begin{equation}
    \vec{V}^{\odot 2} \; \leftarrow \; C_2^{-1} \cdot
    \left(  
    \vec{\phi}_\text{target}
    - \vec{c}_0
    - \sum_{k \neq 2} C_k \cdot \vec{V}^{\odot k}
    \right)
\end{equation}

Before applying a component-wise square root on $\vec{V}^{\odot 2}$, the algorithm checks that all components are positive and below the safe voltage bound. Else, $2\pi$ terms are subtracted or added to the corresponding components of $\vec{\phi}$ until all bound conditions are met.

The solver exits the iterative approximation process when all components of $\vec{\phi}\left(\vec{V}\right)$ are close to $\vec{\phi}_\text{target}$ within some specified accuracy value, typically 0.1 mrad. Note that when the phase relation contains only the order 2 term, the solver delivers the solution in a single iteration.

The described solver bears two major improvements with respect to the one presented in \cite{Fyrillas2024b}:

\begin{itemize}
    \item our solver is compatible with rectangular crosstalk matrices. In that case, $C_2^{-1}$ is the Moore-Penrose pseudo-inverse. The solution $\vec{\phi}\left(\vec{V}\right)$ may however not approximate $\vec{\phi}_\text{target}$ within the specified accuracy value.
    \item the solver scales with $O(\NPS^2)$ (in comparison to the $O(\NPS^3)$ scaling of \cite{Fyrillas2024b}). It is indeed sufficient to compute $C_2^{-1}$ only once before operating the PIC. Each iteration consists then of $O(\NPS^2)$ operations due to matrix-vector multiplication. In addition, the number of iterations needed to approximate a solution is empirically independent of the number of phase shifters. 
\end{itemize}

\section{CROSSTALK MATRIX REDUCTION THEOREMS}
\label{app:reduction_theorems}

The number of phase shifters (PSs) in a photonic integrated circuit (PIC) is denoted $\NPS$ (controlled and induced) and the number of controlled PSs is $\NCPS$. $\phi_k$ denotes the phase shift value implemented by the PS labeled $k$.

\subsection{Lemma for \ref{proof:reduction}}
\label{proof:lemma}
\paragraph{Statement}

Consider a PIC with PSs labeled $1, ..., \NPS$ (including controlled and induced phase shifters) and extended crosstalk matrix $C$ of size $\NPS\times\NCPS$.  

\medskip

Suppose that there exists an invariant phase transformation (see Section \ref{subsec:matrix_transformations}) of the form
\begin{equation}
    \forall i\in \llbracket 1, \NPS \rrbracket, \hspace{0.4cm} \phi_i \longrightarrow \phi_i + \alpha_i \psi
\end{equation}
that holds for all values of $\psi\in[0, 2\pi[$. The coefficients $\alpha_1,...,\alpha_n$ are in $\mathds{R}$.

\medskip

Then the following transformation acting on $C$ is an invariant crosstalk matrix transformation:
\begin{equation}
    \forall i\in \llbracket 1, \NPS \rrbracket, \forall j\in \llbracket 1, \NCPS \rrbracket, \hspace{0.4cm} C_{ij} \longrightarrow C_{ij} + \alpha_i \deltacj
\end{equation}
with $\deltacj\in\mathds{R}$ for all $j\in \llbracket 1, \NCPS \rrbracket$.

\bigskip

\paragraph{Proof}

The extended phase-voltage relation of the PIC is $\PHIEXT = C \cdot \vec{V}^{\odot 2} + \CZEROEXT$ as stated in Eq.~\ref{eq:phase_voltage_extended}. Each phase shift is then expressed as
\begin{equation}
    \phi_i = \sum_{j=1}^{\NCPS} C_{ij} V_j^2 + (c_0)_i
\end{equation}

We apply the matrix transformation specified in the above statement to the crosstalk matrix $C$. The phase shifts evolve according to $\phi_i \longrightarrow \phi_i + \Delta \phi_i$, with 
\begin{equation}
    \Delta \phi_i = \alpha_i \sum_{j=1}^{\NCPS} \deltacj V_j^2
\end{equation}
Note that the phase shift variation is of the form $\Delta \phi_i=\alpha_i \psi$, which defines an invariant phase transformation by assumption. Thus we have proven that the stated crosstalk matrix transformation is invariant. $\qedsymbol$

\subsection{Proof of Theorem \ref{thm:reduction} (Crosstalk matrix row deletion)}
\label{proof:reduction}
\paragraph{Statement}

Consider a PIC with PSs labeled $1, ..., \NPS$ (including controlled and induced phase shifters) and extended crosstalk matrix $C$ of size $\NPS\times\NCPS$. 

\medskip

Suppose that PS $k$ is an induced PS and that there exists an invariant phase transformation of the form
\begin{equation}
    \forall i\in \llbracket 1, \NPS \rrbracket, \hspace{0.4cm} \phi_i \rightarrow
    \begin{cases}
    \phi_i + \alpha_i \phi_k  \hspace{0.2cm} \text{for} \hspace{0.2cm} i \neq k\\
    0  \hspace{1.45cm} \text{for} \hspace{0.2cm} i = k
    \end{cases}
\end{equation}
that holds for all values of $\phi_k\in[0, 2\pi[$. The coefficients $\alpha_1,...,\alpha_n$ are in $\mathds{R}$.

\medskip

Then we can apply the following invariant transformation to the crosstalk matrix $C$, where $R_i$ designates the i-th row of $C$:
\begin{equation}
    \forall i\in \llbracket 1, \NPS \rrbracket, \hspace{0.4cm} R_i \rightarrow
    \begin{cases}
    R_i + \alpha_i R_k  \hspace{0.2cm} \text{for} \hspace{0.2cm} i \neq k\\
    0  \hspace{1.45cm} \text{for} \hspace{0.2cm} i = k
    \end{cases}
\end{equation}
The row $R_k$ is then deleted. 
\\

\paragraph{Proof} 

The stated invariant transformation can be condensed into
\begin{equation}
    \forall i\in \llbracket 1, \NPS \rrbracket, \hspace{0.4cm} \phi_i \rightarrow
    \phi_i + \alpha_i \phi_k  
\end{equation}
by setting $\alpha_k = -1$. Let us focus on a single column $j$ of the crosstalk matrix $C$. By Lemma \ref{proof:lemma}, the following transformation is an invariant crosstalk matrix transformation:
\begin{equation}
    \forall i\in \llbracket 1, \NPS \rrbracket, \hspace{0.4cm} C_{ij} \rightarrow C_{ij} + \alpha_i \deltacj
\end{equation}
with $\deltacj\in\mathbb{R}$. Taking $\deltacj = C_{kj}$, we have
\begin{equation}
    \forall i\in \llbracket 1, \NPS \rrbracket, \hspace{0.4cm} C_{ij} \rightarrow
    \begin{cases}
    C_{ij} + \alpha_i C_{kj}  \hspace{0.2cm} \text{for} \hspace{0.2cm} i \neq k\\
    0  \hspace{1.45cm} \text{for} \hspace{0.2cm} i = k
    \end{cases}
    ,
\end{equation}
Note that the matrix element of column $j$ corresponding to $\phi_k$ has been zeroed. Applying the same procedure for every column of $C$, we zero the entire row $R_k$ and retrieve the stated transformation on the rows of $C$ in the statement of the theorem. Because the row corresponding to $\phi_k$ is zero everywhere, the phase shift value is then always zero, hence we can remove the induced PS $k$ from the circuit and delete the associated row in the matrix. $\qedsymbol$

\section{PHASE SIMPLIFICATION ALGORITHM}
\label{app:phase_algo}

\subsection{Algorithm}

We introduce a phase simplification algorithm to efficiently find the removable induced phase shifters (PSs) in a given photonic integrated circuit (PIC) and the associated invariant phase transformations (see Section \ref{subsec:circuit_rewriting}). We say that an induced PS is \textit{removable} when its implemented phase shift can be moved out of the circuit as exemplified in Fig.~\ref{fig:rewriting}c. This yields an invariant phase transformation of the form Eq.~\ref{eq:pit} that enables the removal of the induced PSs from the circuit, following Theorem \ref{thm:reduction}.

An induced PS is removable if and only if it is adjacent (see Section \ref{subsec:circuit_rewriting}) to three components of the types
\begin{itemize}
    \item phase-invariant input/output port,
    \item controlled PS,
    \item removable induced PS.
\end{itemize}
We can then indeed apply the rule $\phi$-remove (see Fig.~\ref{fig:rewriting}b) to propagate the phase shift out of the circuit. Note that as the matrix reduction procedure progresses, some removable induced PSs may become \textit{non-removable} due to the gradual deletion of induced PSs in the circuit. 

One approach to finding the removable induced PSs of a given circuit would be to start from each induced PSs and use a recursive function to determine if the adjacent induced PSs are removable themselves. Our phase simplification algorithm does the opposite by starting from the induced PSs that are trivially removable, i.e.\ adjacent to three controlled PSs or phase-invariant input/output ports. These trivially removable induced PSs are labeled as removable by the algorithm and the associated linear reduction coefficients ($\alpha_k$ in Fig.~\ref{fig:reduction}a) are stored. The algorithm subsequently processes in the same way induced PSs adjacent to the newly labeled induced PSs. The procedure is iterated until no new adjacent removable induced PSs can be found.

If there are however still induced PSs left in the circuit at this stage, this implies that the circuit features at least one non-removable induced PS. In that case, the algorithm designates one of the remaining unlabeled induced PSs as non-removable. The designated PS must be adjacent to the already labeled ones. Non-removable induced PSs are on equal footing with controlled PSs in the sense that they are static in the circuit and can absorb adjacent phase shifts. The algorithm then resumes its iterative procedure until all induced PSs have been labeled. We prove in App.~\ref{proof:maximal} that the number of non-removable induced PSs found by our phase simplification algorithm is minimal.

The presence of a non-removable induced PS in a circuit entails that its initial rectangular crosstalk matrix cannot be reduced to a square matrix by linear transformations by repeated application of Theorem \ref{thm:reduction}. As shown on Fig.~\ref{fig:reduction}d, this does not provide an optimal PIC control accuracy but improves nevertheless the control accuracy of the circuit. 

The presented phase simplification algorithm is flexible as it can be extended with additional rules for removing induced PSs if the circuit features components other than phase shifters and beamsplitters. This would just change the definition of a removable induced PS.

\begin{figure*}[ht]
    \centering
    \includegraphics[width=0.4\linewidth]{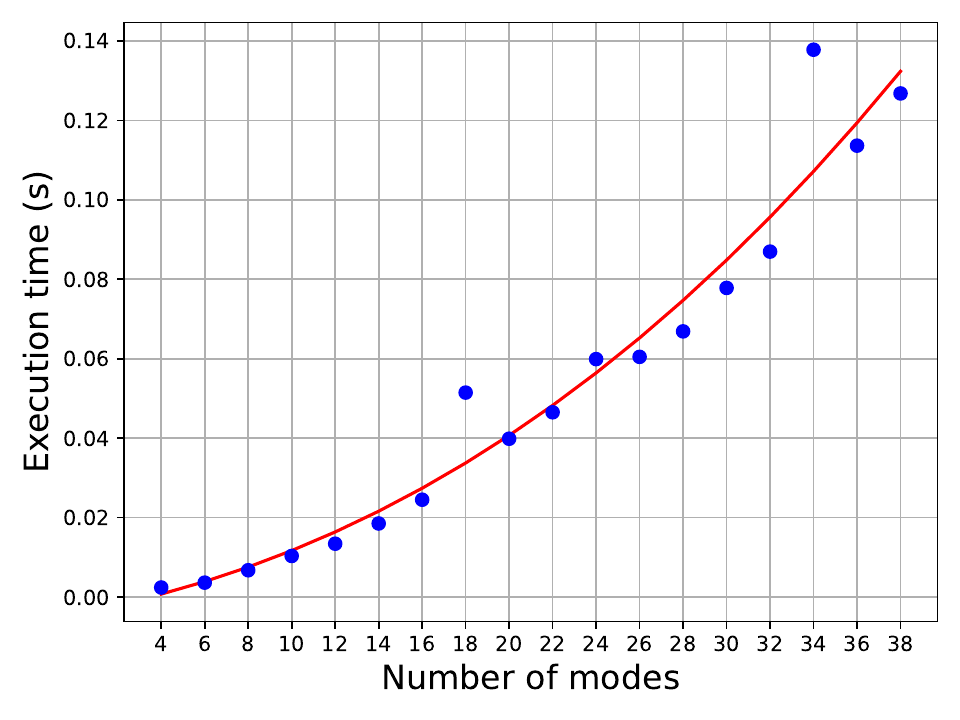}
    \caption{
    \textbf{Reduction algorithm evaluation on Clements interferometers}.
    The algorithm is implemented in Python. Each blue point consists in 10 repetitions of the algorithm. We only benchmark the algorithm that establishes the linear relations between the phase shifters, not the crosstalk matrix transformation process. The red curve is a third degree polynomial fit. 
    }
    \label{fig:reduction_algo_benchmark}
\end{figure*}

\subsection{Algorithmic complexity}

We now discuss the complexity of the phase simplification algorithm. We consider first the general case. The total number of induced PSs to label is less than $\NPS$, as we never reconsider already labeled induced PSs. Establishing a phase relationship between PSs is at most of complexity $\NPS$ because a relation could involve all the on-chip PSs. Therefore, the complexity of the full procedure is $O(\NPS^2)$ in the worst case.

In our case, the linear phase relationships are local, thus the relations do not always involve all the on-chip PSs. Thus the stated upper bound is not attained in practice. The reduction algorithm is benchmarked on Clements interferometers with the execution time graphed on Fig.~\ref{fig:reduction_algo_benchmark}. We observe that the algorithm scales with $O(m^3)$ with $m$ the number of modes. In a Clements interferometer, the number of phase shifters $\NPS$ is related to $m$ via $m\approx \sqrt{\NPS}$. Hence, the scaling of the algorithm is $O(\NPS^{3/2})$ in our case.

\section{PROOF OF CROSSTALK-ROBUSTNESS CRITERION}

We prove the graphical criterion of Section \ref{subsec:criteria} that certifies the robustness of a given photonic integrated circuit (PIC) with respect to crosstalk.

\subsection{Proof of Theorem \ref{thm:robust} (Crosstalk-robustness criterion)}
\label{proof:robust}
\paragraph{Statement}

A PIC is crosstalk-robust if and only if its associated pruned graph is acyclic.

\bigskip

\paragraph{Proof}

\begin{figure*}[ht]
    \centering
    \includegraphics[width=\linewidth]{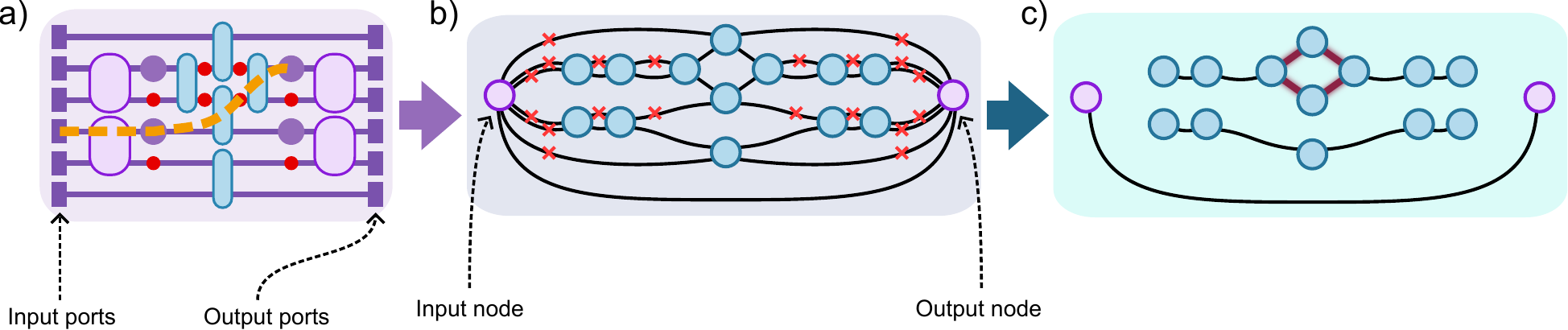}
    \caption{
    Graph representation of Fig.~\ref{fig:cycles}b. Red edges indicate phase invariant input/output ports. Edges corresponding to waveguides featuring a controlled PS are marked with a red cross.
    }
    \label{fig:proof_illustration}
\end{figure*}

Let us consider an arbitrary PIC, such as the one of Fig.~\ref{fig:proof_illustration}a. We call \textit{path} a sequence of phase shifters (PSs) that extends from one PS/input port/output port to another PS/input port/output port (see orange path on Fig.~\ref{fig:proof_illustration}a). 

Let us consider an induced PS in the PIC. From the rule $\phi$-remove (see Fig.~\ref{fig:rewriting}b) and App. \ref{app:phase_algo}, the induced PS is removable when it is adjacent to three components of the following type
\begin{itemize}
    \item phase invariant input/output port,
    \item controlled PS,
    \item removable PS.
\end{itemize}
By this observation, an induced PS is removable to the left (resp. right) when all paths starting from the left (resp. right) of the induced PS eventually reach a phase invariant input/output port or a controlled PS. An induced PS is removable when it is removable to the left or to the right. Consequently, an induced PS is non-removable when it is non-removable to the left and to the right. In other words, an induced PS is non-removable when one of the paths to the left and to the right ends on a phase-dependent input/output port or performs a cycle.

We now convert the PIC into a mathematical graph. The nodes represent the on-chip beamsplitters and the edges interconnecting the nodes symbolize the PIC waveguides. The edges corresponding to PIC waveguide section featuring a controlled PS are marked with a cross. The graph features two additional nodes, called \textit{input} and \textit{output} nodes. Every input (resp. output) port of the circuit is connected to the input (resp. output) node, marked with a cross if the port is phase invariant. The input and output nodes are connected by an edge as well. Any path in the PIC has a corresponding path in the graph. 

We obtain a \textit{pruned graph} by deleting every edge marked with a cross. Internal cycles in the pruned graph (connecting beamsplitter nodes) mark cyclic paths in the PIC consisting only of induced PSs, which is the signature of the presence of a non-removable induced PS. 

Note that if an induced PS in the PIC has paths to the left and to the right leading to phase-dependent input/output ports, it is non-removable. As the input and output nodes are connected in the graph, this non-removability also manifests as the presence of a cycle.

This proves the theorem. $\qedsymbol$

\subsection{Proof of maximal reduction for non-crosstalk-robust interferometers: circuit rank of the pruned graphs}
\label{proof:maximal}

When the interferometer is not crosstalk-robust, then by definition its crosstalk matrix cannot be fully reduced to a square matrix (see Section \ref{subsec:criteria}). Non crosstalk-robust interferometers feature induced PSs that cannot be removed from the circuit. We prove here that the number of removed induced PSs following our phase-simplification algorithm of App.~\ref{app:phase_algo} is maximal. Equivalently, this shows that the number of remaining rows in the reduced crosstalk matrix is minimal.

We use the pruned graph formalism introduced in Section \ref{subsec:criteria}. As stated by Theorem~\ref{thm:robust} and shown in App.~\ref{proof:robust}, an induced PS cannot be removed if its corresponding edge in the pruned graph belongs to a cycle (see Fig.~\ref{fig:proof_illustration}). Reciprocally, if the pruned graph is acyclic, then every induced PS can be removed from the circuit. The remaining edges in the pruned graph represent the induced PSs of the circuit. Therefore, the minimal number of non-removable PSs is the minimal number of edges to remove in the pruned graph to make it acyclic. This number is known as the circuit rank \cite{berge2001}, leading to the following theorem:
\begin{theorem}[Minimal number of non-removable PS]
    The minimal number of non-removable PS in a given PIC is the circuit rank of its associated pruned graph.
\end{theorem}
As the pruned graph is undirected, the circuit rank $r$ is straightforward to compute from 
\begin{equation}
    r=e-v+c,
\end{equation}
where $e$ is the number of edges, $v$ is the number of vertices, and $c$ is the number of connected components in the pruned graph. For instance in Fig.~\ref{fig:proof_illustration}, we have $e=13$, $v=15$ and $c=3$, yielding indeed $r=1$, i.e.\ a single edge must be removed from graph to make it acyclic.

Remarkably, it is possible to find a minimum set of edges breaking all the pruned graph cycles using a greedy algorithm \cite{berge1973graphs}. At each step, the greedy algorithm chooses to cut an edge belonging to at least one cycle of the pruned graph, and continues until there is no cycle. A cut edge by the algorithm in the pruned graph corresponds, in the phase simplification algorithm, to label a non-removable induced PS. 
More precisely, our phase-simplification algorithm of App.~\ref{app:phase_algo} follows the greedy approach: when no additional induced PS can be identified as removable, the algorithm randomly selects one of the remaining unlabeled induced PS and designates it as non-removable (cutting a random edge in a cycle) and continues until every PS is removable (until the pruned graph is acyclic). The greedy algorithm being optimal for finding the minimum set of edges breaking all the cycles, the phase simplification algorithm gives the minimum number of non-removable PSs, thus the reduction is maximal.

\section{CROSSTALK-ROBUSTNESS CRITERION WITH PHASE-DEPENDENT INPUT PORTS}
\label{app:phase_dependent}

\begin{figure*}[ht]
    \centering
    \includegraphics[width=\linewidth]{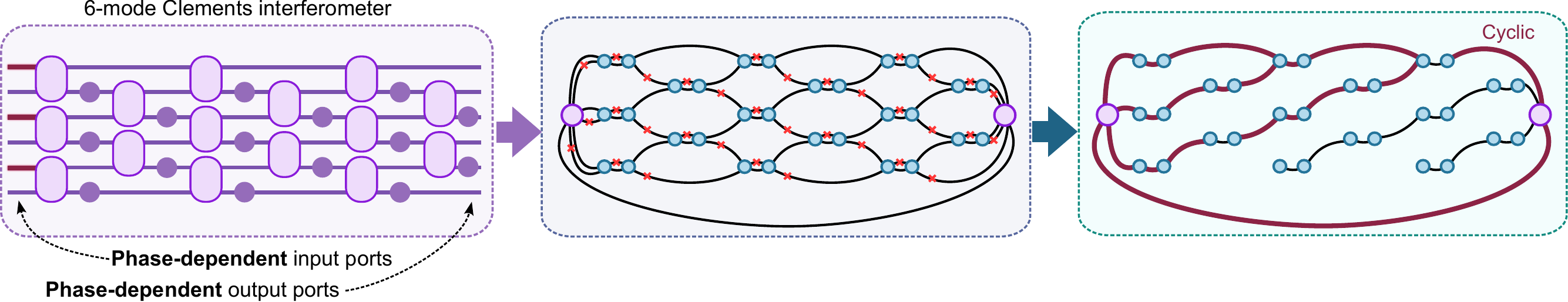}
    \caption{
    \textbf{Applying the crosstalk-robustness criterion for an interferometer with phase-dependent input ports}.
    }
    \label{fig:criterion_phase_dependent}
\end{figure*}

We illustrate in Fig.~\ref{fig:criterion_phase_dependent} the crosstalk-robustness criterion of Section \ref{subsec:criteria} for an interferometer with phase-dependent (opposite of phase-invariant) input and output ports. In practice, this is the case when using the interferometer with coherent states of light and the output lead to a second PIC for instance. The considered interferometer is a variant of the 6-mode Clements interferometer with controlled phase shifters (PSs) on some output ports. Applying the crosstalk-robustness criterion reveals that in this configuration, the Clements interferometer is vulnerable to crosstalk, because its pruned graph contains cycles. Adding controlled PSs outside of the MZI mesh, such as in the original Clements mesh \cite{Clements2016}, restores the crosstalk-robustness of the scheme.

\section{CONVERGENCE COMPARISON WITH BELL-WALMSLEY INTERFEROMETER}
\label{app:bell_comparison}

\begin{figure*}[ht]
    \centering
    \includegraphics[width=0.5\linewidth]{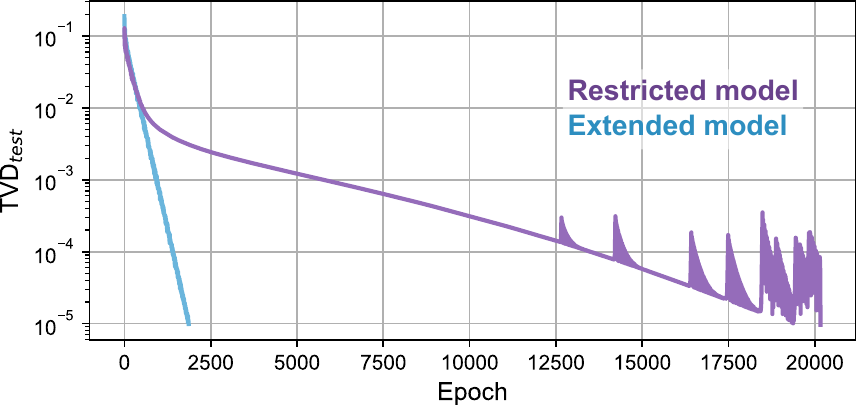}
    \caption{
    $\text{TVD}_\text{test}$ as a function of the number of epochs processed for characterizing a 12-mode Bell interferometer. The purple (resp.\ purple) curve is associated to a machine learning model equipped with a restricted (resp.\ extended) crosstalk model.
    }
    \label{fig:bell_comp}
\end{figure*}

Similarly to Section \ref{sec:charac}, we train machine learning models (MLMs) equipped with restricted and extended crosstalk models to characterize a 12-mode Bell-Walmsley interferometer \cite{Bell2021}. Fig.~\ref{fig:bell_comp} shows that the extended MLM converges in substantially fewer epochs to the $\text{TVD}_\text{test}$ target of $10^{-5}$ than the restricted MLM. The characterization duration is 6.5 minutes for the extended MLM and 53.3 minutes for the restricted MLM. All the simulation and training parameters are the same as described in the Methods section, but the simulation was performed with a different CPU and GPU. The ripples in the purple curve in Fig.~\ref{fig:bell_comp} can be removed by decreasing the learning rate of the model, at the cost of slower descent.

\end{document}